\documentclass[5p,twocolumn,authoryear]{elsarticle}

\usepackage{amssymb}
\usepackage{amsmath}

\usepackage{lineno}

\usepackage[utf8]{inputenc}
\usepackage[T1]{fontenc}
\usepackage{textcomp}
\usepackage{graphicx,multirow,url,color,xspace}

\usepackage{rotating}
\usepackage{subfigure}
\usepackage{hyperref}

\usepackage{txfonts}

\newcommand{\msun}{{M}_{\odot}}

\newbox\grsign \setbox\grsign=\hbox{$>$} \newdimen\grdimen \grdimen=\ht\grsign
\newbox\simpropbox
\setbox\simpropbox=\hbox{\raise.5ex\hbox{$\propto$}\llap
 {\lower.5ex\hbox{$\sim$}}}\ht2=\grdimen\dp2=0pt
\def\simprop{\mathrel{\copy\simpropbox}}

\journal{New Astronomy Review}

\begin{document}

\begin{frontmatter}

\title{Spins of Black Holes in X-ray Binaries and the Tension with the Gravitational Wave Measurements}

\author[inst1]
{Andrzej A. Zdziarski \corref{cor}}
\affiliation[inst1]{{Nicolaus Copernicus Astronomical Center,}, {Polish Academy of Sciences,}, {Bartycka 18,}, {PL-00-716 Warszawa,}, Poland}

\cortext[cor]{Corresponding author:  aaz@camk.edu.pl}

\author[inst2]{Gr{\'e}goire Marcel}
\affiliation[inst2]{{Department of Physics and Astronomy,}, {FI-20014 University of Turku,}, Finland}

\author[inst2,inst3]{Alexandra Veledina}
\affiliation[inst3]{{Nordita,}, {KTH Royal Institute of Technology and Stockholm University,}, {Roslagstullsbacken 23,}, {SE-10691 Stockholm,}, Sweden}

\author[inst4]{Aleksandra Olejak}
\affiliation[inst4]{organization={Max Planck Institut f{\"u}r Astrophysik}, addressline={Karl-Schwarzschild-Stra{\ss}e 1}, {85748 Garching,}, Germany}

\author[inst1,inst5]{Debora Lan{\v{c}}ov{\'a}}
\affiliation[inst5]{organization={Research Center for Computational Physics and Data Processing, Institute of Physics, Silesian University in Opava}, addressline={Bezru\v{c}ovo n\'{a}m. 13, 746 01 Opava}, Czech Republic}

\begin{abstract}
We review current challenges in understanding the values and origins of the spins of black holes in binaries. Thanks to recent advances in astrophysical instrumentation, the spins can now be measured using both gravitational waves emitted by merging black holes and electromagnetic radiation from accreting X-ray binaries containing black holes. A key finding of the gravitational-wave observatories is that premerger black holes in binaries have low spin values, with an average dimensionless spin parameter of $a_*\sim$0.1--0.2, with 90\% having $a_*\lesssim 0.6$. This implies that the natal spins of black holes are generally low, and the angular momentum transport in massive stars is efficient. On the other hand, most of the published spins in X-ray binaries are very high, $a_*\gtrsim 0.7$. In particular, this is the case for binaries with high-mass donors (potential progenitors of mergers), where their published spins range from 0.8 to 1.0. At the same time, their short lifetimes prevent significant spin-up by accretion. Those with low-mass donors could be spun-up to $a_*\gtrsim 0.5$ by conservative accretion. Spins $a_*\gtrsim 0.7$ can be achieved only if the donor initial masses were more than several solar masses, which remains unproven. However, the existing methods of spin measurements suffer from significant systematic errors. The method relying on relativistic X-ray line broadening is based on the separation of the observed spectra into incident and reflected ones, which is intrinsically highly uncertain. The method relying on spectral fitting of accretion disk continua uses models that predict the disk to be highly unstable, while stability is observed. Improved stable models yield disk temperatures higher than the standard models, and consequently predict lower spins. The published spin measurements in X-ray binaries are uncertain. The spins of the binaries with high-mass donors may be low, while those with low-mass donors have a broader spin distribution, ranging from low to high, including high spins as required to power relativistic jets.


\end{abstract}





\end{frontmatter}




\section{Introduction}

Black holes (BHs) are the simplest objects in the universe, fully characterized by only three fundamental parameters: mass, spin, and charge. Then, astrophysical BHs would quickly discharge into their surroundings \citep{Komissarov22}, leaving us with mass and spin. The masses of stellar BHs in binaries can be reliably measured by studying both gravitational waves (GWs) from merging binary BHs (BBHs) and electromagnetic (EM) radiation from accreting BH X-ray binaries (XRBs) using the radial velocities of their donors. The ranges of the masses of both kinds of BHs overlap, and those from GWs extended to higher values (up to $\gtrsim 100\msun$) than those found in XRBs ($\lesssim 20\msun$), which appears to be well explained (see Section \ref{GW}). 

The opposite situation occurs for the spins, where the measured spins of premerger BHs are very low on average, while those of BHs measured in XRBs are high on average and, in many cases, close to the maximum value. In particular, the measured spins of BHs in XRBs with high mass donors (HMXBs), which can be progenitors of BBHs, are very high, in stark contrast to the case of BBHs. The origin of this tension remains unclear. 

The values of the spins are critically important for our understanding of stellar evolution, BH formation scenarios, accretion, and testing theories of relativity. For accretion, they are essential for its efficiency, the disk emission, and the launching of jets. Furthermore, high spin values are crucial for understanding the origin of gamma-ray bursts (GRBs). In particular, if a rapidly rotating star collapses into a BH, the resulting accretion disk can launch a jet thought to power long-duration GRBs \citep{BZ77}.

Here, we review the observational results from both GW and EM emissions and the methods of spin measurements. We then search for possible ways to resolve the spin tension. We find that while the spins determined by the GWs appear highly reliable, the methods of measuring the spins by the EM emission are not. The method using disk reflection suffers from the uncertainty of separating the reflected emission from the incident one. The method using relativistic disk continuum emission suffers from the instabilities predicted by analytical disk models, while stability is observed. This renders them not applicable to describe the emission of BH XRBs, and calls for the development of new stable models suitable for spectral fitting.

We use the standard dimensionless spin parameter, 
\begin{equation}
a_*\equiv J c/G M^2,
\end{equation}
where $J$ and $M$ are the angular momentum and mass, respectively, and $G$ and $c$ are the gravitational constant and the speed of light. The BH rotation period at the horizon as seen at infinity is (e.g., \citealt{Lightman75})
\begin{equation}
P=\frac{4 \pi G M(1+\sqrt{1-a_*^2})}{a_* c^3}.
\label{period}
\end{equation}

In Sections \ref{GW} and \ref{EM}, we review the spins from BH mergers and studies of BH XRBs, respectively. Within the latter, Sections \ref{spectral}--\ref{polarization} and \ref{published} review the methods of measuring the spins and the published values of the spins, respectively. Section \ref{solutions} is devoted to possible resolutions of the tension, where \ref{revised} discusses possibilities of revising the EM measurements, \ref{accretion} covers the feasibility of accretion to spin up BHs, and \ref{population} discusses the possibility that BH XRBs and BBHs belong to different stellar populations. We discuss and summarize our results in Section \ref{discussion}. 

\section{Results from BH mergers}
\label{GW}

Mergers of BBHs can now be studied using GWs, which have allowed us to measure the spin of the premerger BHs. At least some BBHs (excluding those dynamically formed) are the final stages of binaries containing a BH and a high-mass donor. The latter includes currently accreting BHs, i.e., BH HMXBs. The results of these studies thus have important implications for comparing the populations of BBHs and BH XRBs, as well as for understanding their differences and similarities.

Following the first detection of GWs from the merger of two BHs in 2015 \citep{Abbott16}, more than 200 such events at redshifts $\lesssim$1 have so far been publicly announced by the LIGO-Virgo-KAGRA collaboration (the recent GWTC-4.0 catalog by \citealt{LVK25a}  now contains 218 candidates with probabilities $\geq 0.5$). Two main results about the properties of the premerger BHs have been obtained.

One is that the distribution of the measured BH component masses in BBHs extends to much higher values, up to $\gtrsim\! 100\msun$ \citep{GW2311232}, than those observed in accreting BH XRBs. Until recently, the highest measured BH mass in XRBs was that of Cyg X-1, $21.2\pm 2.2 \msun$ \citep{Miller-Jones21}. However, the recent investigation using detailed atmospheric (non-LTE) models of the donor by \citet{Ramachandran25} found $M\approx (12.7$--$17.8)\msun$. We also note that a $33\pm 1 \msun$ BH was recently discovered in a detached binary, Gaia BH3, within the Galaxy \citep{Gaia24}. However, as discussed in that work, the donor has a very low metallicity (allowing the BH progenitor to avoid strong mass loss in stellar winds), resulting in significantly higher expected BH masses in the halo compared to other Galactic components \citep{Olejak2020}. The system is associated with the metal-poor ED-2 stream, suggesting the role of dynamical interactions in its formation \citep{Pina24}. The mass range difference between EM and GW populations of BHs can be explained by both the observational selection effects, with mergers of heavier BHs giving stronger signals in GWs \citep{Fishbach22, Liotine23}, and the effect of the metallicity decreasing with the increasing redshift, which then reduces the rates of stellar winds, e.g., \citet{Belczynski20}. 

The other major result is that the spins of the premerger BHs are generally low. The population study of \citet{LVK25b} shows that the effective spin parameter, which is the average of the individual spin parameters, $a_{1,2}$, weighted by the masses projected onto the orbital angular momentum, 
\begin{equation}
a_{\rm eff}\equiv \frac{M_1 a_1 \cos\theta_1+M_2 a_2 \cos\theta_2}{M_1+M_2},
\label{aeff}
\end{equation}
peaks at $a_{\rm eff} \approx 0.0$, with the inferred distribution skewed toward positive values extending only up to $\approx$0.5 (see figure 9 in \citealt{LVK25b}). Here, $M_{1,2}$ are the masses of the two BHs, and $\theta_{1,2}$ are the angles of the respective spin axes with respect to the binary orbital momentum axis. The leading spin-orbit coupling term determines the value of the effective spin in the post-Newtonian waveform fitted to the gravitational signal. Its values are measured with better precision than those of the individual BHs. The modeled distribution of the individual spins peaks within $a_*\approx 0.01$--$0.23$ (at 90\% credibility), and $\sim$90\% of them are $a_*\lesssim$0.57 \citep{LVK25b}. 

\begin{figure}
\centerline{\includegraphics[width=8cm]{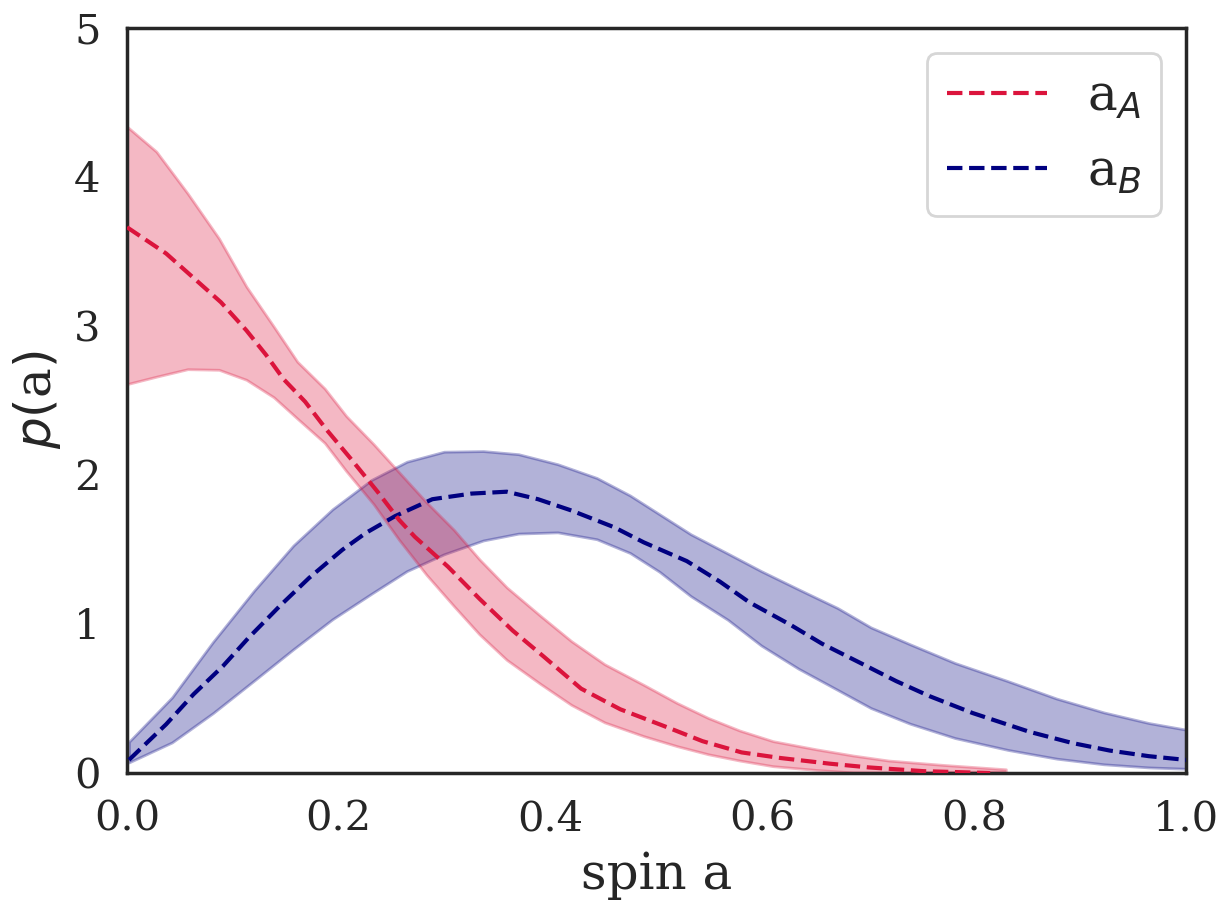}}
    \caption{Distribution of the individual values of the lower (denoted by $a_A$) and higher ($a_B$) spins of the components of premerger BH binaries. Adapted from \citet{LVK25b}. See Section \ref{GW} for discussion. 
} 
\label{fig17}
\end{figure}

\begin{figure}[b!]
\centerline{\includegraphics[width=\columnwidth]{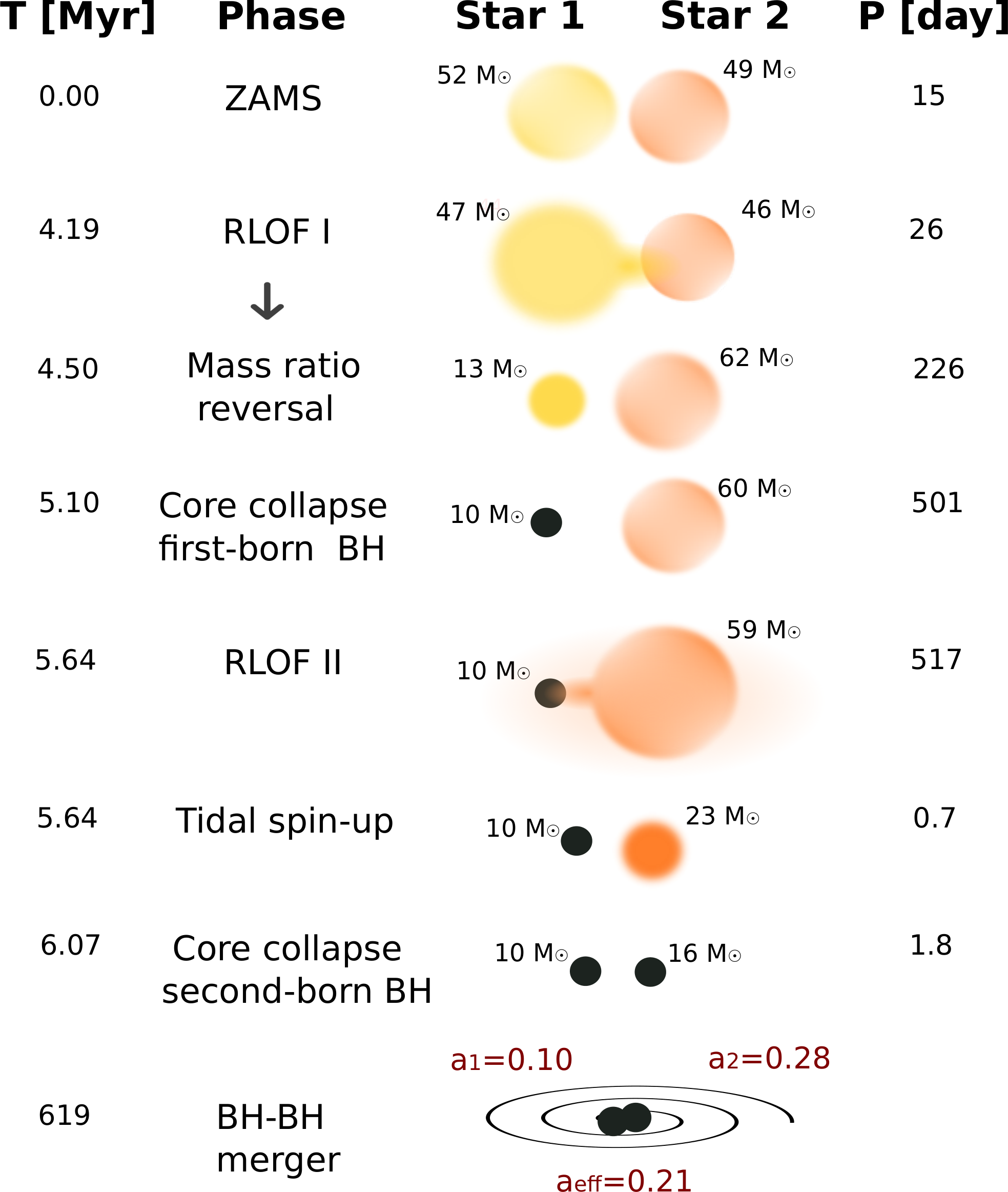}}
    \caption{An example of an evolutionary scenario leading to the formation of a merging BBH system that involves a mass ratio reversal and tidal spin-up of the second-born BH. Based on the example described by \citet{Olejak24}. See Section \ref{GW} for a description of the evolution states. The initial eccentricity at the zero-age main sequence (ZAMS) is $e=0.40$, but the orbit is circularized quickly already in RLOF I.}
\label{evolution}
\end{figure}

Furthermore, the distributions of the higher and lower spins of the BHs in the premerger binaries differ significantly (as first noted by \citealt{Biscoveanu21}). This is shown in Fig.\ \ref{fig17} \citep{LVK25b}, where we see that their distributions peak at $a_A= 0.0$ and $a_{B}\approx 0.3$--0.4, respectively. A possible explanation for that is that the first-born BH rotates very slowly (at its low natal spin, see below), and then some process after its formation spins up the other BH.  The scenario proposed by recent literature involves tidal interactions of the first-born BH with a stripped progenitor of the other BH in a tight orbit, $\lesssim$1 day, which leads to the synchronous rotation of the BH progenitor \citep{Qin18, Mandel20, Ma_Fuller23}. These interactions occur after the formation of the first BH but before the formation of the second one \citep{Olejak21}, as shown in the phase `Tidal spin-up' of Fig.\ \ref{evolution}.

The low natal spins of BHs are explained by the stellar core during expansion (occurring when leaving the main sequence) remaining strongly coupled to the outer envelope, as predicted by standard stellar models with efficient angular momentum transport \citep{Spruit02, Fuller_Ma19, Belczynski20}. Those models yield $a_*\sim 0.01$--0.1. The presence of efficient angular momentum transfer (and thus low natal spins) is also supported by asteroseismology constraints \citep{Benomar15, Gehan18, Aertes2019}, which are, however, mostly limited to low-mass stars, with some works extending into the range of NS progenitors \citep{Burssens2023}.  The opposite situation, with the first-born BH fast spinning (at a high natal spin), would require an efficient spin-down process for the other BH, which is difficult to invent. 

Moreover, there is also evidence that some of the slower-spinning BHs (with $a_{A}$) are also less massive ones, contrary to the expectation that the more massive BH should be formed first (due to the evolutionary timescales). In the sample of \citet{Abbott23}, 37 BBHs have the more massive BH with a lower spin, 22 of them have the less massive BH with a lower spin, and 6 BBHs have both BHs with the same best-fit spin. These numbers are, however, subject to substantial statistical and systematic errors \citep{Mould22, Gerosa25}. Still, such data provide evidence for the presence of some slower and less massive premerger BHs in the sample observed so far.

A promising scenario to make the less massive BH the slower spinning one involves mass ratio reversal during the first stable mass transfer, i.e., before the BH formation \citep{Olejak21, Broekgaarden22, Adamcewicz24, Olejak24}. Such a scenario is illustrated in Fig.\ \ref{evolution}, adapted from \citet{Olejak24}. In it, Star 1 is heavier on the zero-age main sequence (ZAMS), and it efficiently transfers most of its mass (in its hydrogen envelope) to the companion Star 2 via Roche-lobe overflow (RLOF I). After that, it becomes a stripped helium-core star (Wolf-Rayet; WR) and the lighter of the two, while the binary becomes detached. Then, the WR star collapses to a BH with a low (natal) spin and with a much heavier companion. Next, the Star 2 evolves and starts to transfer mass to the BH companion (RLOF II). The accretion on it is, however, Eddington-limited. Therefore, it only slightly increases the BH mass during that short evolution stage, with most of the transferred mass of Star 2 being lost to outflows. Star 2 becomes then a WR star. This second stable mass transfer phase leads to a significant reduction of the binary separation due to a highly unequal mass ratio with the donor significantly more massive than the BH \citep[see orbital evolution during mass transfer phase, e.g., in][]{Tauris23}. The binary period is now so short that the tidal forces exerted by the BH on the companion spin it up, leading to its corotation with the binary. When Star 2 collapses to a BH, it has both the mass and the spin higher than the first-born BH. Finally, both BHs merge.

A likely candidate for a system that underwent a tidal spin-up phase is the HMXB Cyg X-3. While the nature of its compact object is not certain, it is most likely a BH with the mass estimated as $\approx 7 M_{\odot}$, while the donor is a WR star with the mass estimated as $\approx 12 M_{\odot}$, and they orbit each other at the period as short as 0.20 d (\citealt{Antokhin22} and references therein). The current number of similar type X-ray binary systems with WR donors that could later evolve into a BBH merger in the Milky Way is expected to be very low due to the high metallicity of star-forming regions \citep{Sen25}. 

The scenario with the fast-spinning second-born BH via tidal spin-up has also been considered to facilitate the production of long-duration GRBs (see, e.g., \citealt{Fryer22, Bavera22}). The leading model for the production of those GRBs requires a collapse of a rapidly rotating massive star forming an accretion disk that, in turn, launches a strong jet powered by BH spin extraction  \citep{BZ77}. The propagating jet then produces the GRB \citep{Woosley93, MacFadyen99, MacFadyen01}. 

Another possible mechanism that could spin up BHs is accretion. However, it would need to be several orders of magnitude above the Eddington limit to significantly affect the spin and masses of merging BHs \citep{vanSon20}. Such models are disfavorable on physical grounds, see the discussion of hypercritical accretion in the context of BH HMXBs in Section \ref{HMXB}. Most currently used population models either adopt the Eddington limit or exceed it slightly. Still, the impact of hypothetical hypercritical accretion on the distribution of individual spins and properties of a population of GW sources has not yet been well explored. 

Besides the spin magnitude, an important signature of the formation mechanism is the spin orientations. The orientation between individual BH spins and the binary orbit peaks at zero angle, but the angular distribution between the two spin axes covers the full range of $\cos \theta$; see figure 7 in \citet{LVK25b}. Such a distribution and, in particular, anti-alignment between the spins is not expected in the classical isolated binary evolution scenario (although see the recent review by \citealt{Baibshav24}). The misalignment of BBH systems formed in isolation could be a signature of significant natal kicks acquired during the formation of compact objects. Natal kicks, however, are not expected to be very high, especially in the case of most massive BHs \citep{Fryer12, Janka24}. A possible mechanism that could result in more misalignment in binary evolution is the so-called spin tossing during BH formation (in either supernova or core collapse) proposed by \citet{Tauris22}, see also \citet{Larsen25}. However, the detailed physical mechanism behind that remains uncertain. 

The probable cause of spin misalignment inferred for a fraction of GW sources is the contribution of systems formed through dynamical interactions, e.g., in dense stellar environments. Mergers originating from such environments are generally expected to exhibit an isotropic spin distribution, symmetric around an effective spin of $a_{\rm eff} \approx 0$ \citep{Rodriguez18, Arca_Sedda20}. The spin distribution inferred from GW detections shows, however, an asymmetry, with a preference for positive effective spin values \citep{Abbott23, LVK25b}; therefore, it cannot be fully explained by the dynamical channel itself. A scenario in which the population is dominated by isolated binaries, but includes a non-negligible fraction of dynamically formed systems, remains a plausible option. Interestingly, a distribution of effective spins consistent with current GW observations could also arise from mergers occurring in young massive clusters \citep{Banerjee23}. In such environments, a significant fraction of binaries may preserve their initial spin alignment asymmetries, since their orientations are not fully randomized by dynamical interactions.

If the dynamical formation channels contribute to the detected population of BBH mergers, some systems may originate from second-generation mergers, in which at least one of the BHs is itself the remnant of a previous BBH coalescence. Repeated BBH mergers are expected to leave a distinct imprint on the spin of the resulting remnant, introducing an additional layer of complexity when interpreting GW data. For stellar-mass BHs formed through hierarchical mergers, the resulting spin magnitude distribution (determined by the orbital angular momentum of the progenitor binary) tends to peak at high values of around $a_*\approx 0.7$, with magnitudes below $\sim$0.5 and above $\sim 0.8$ being unlikely \citep[e.g.,][]{Hofmann16, Gerosa17, Fishbach17}. 

Recent studies by \citet{Tong25} and \citet{Antonini2025} suggest the presence of a subpopulation of dynamically assembled second-generation mergers in the GW catalog, characterized by higher masses and larger effective spins \citep{LVK25a}. In particular, they argue that BBHs with component masses within the so-called pair-instability mass gap \citep{Bond1984, Heger2002, Fryer2001} involve a BH formed through a previous merger. They claim that such a scenario would place the lower edge of the gap at around 45--$50 \msun$. However, this interpretation faces some challenges, e.g., due to the existence of several events in which both components may reside within the mass gap (such as GW231028) and the observed prevalence of positive effective spins across the population, especially among the most massive systems \citep{LVK25b, LVK25a}. This distribution suggests preferential spin alignment in merging BBH binaries, contrary to expectations for the dynamical channel.

One of the most intriguing events in GWTC-4 is GW231123 \citep{GW2311232}, which represents the most massive BBH merger detected to date. Its both component masses were around $100\,\msun$ each, and the BHs' spins exhibit exceptionally large values of $0.90^{+0.10}_{-0.19}$ and $0.80^{+0.20}_{-0.51}$, standing out from the rest of the population. GW231123 poses significant challenges for classical dynamical hierarchical formation in star clusters. Such extremely high spins are unlikely for pure dynamically assembled systems \citep{Fishbach17, Gerosa17}, and the large recoil kicks threaten the retention of the merger remnant \citep{Antonini2016}. Simulations indicate that only a small fraction of hierarchical mergers can reproduce such high spins since the progenitor BHs require well-aligned spins. This suggests that GW231123-like events may require stellar binary evolution either alone \citep{Croon2025, Popa2025, Gottlieb2025, Tanikawa2025}, or with subsequent dynamical assembly \citep{Stegmann2025}. Alternatively, they might form in more exotic environments, such as active galactic nucleus (AGN) disks \citep{Delfavero2025, Bartos2026, Kirouglu2025}.

We note that the interpretation of GW observations is subject to systematic biases and methodological limitations, including those arising from the restricted set of available waveform templates \citep{Divyajyoti2024, Dhani25}. Although the inferred fractions and detailed shape of the underlying spin distribution depend on the chosen modeling framework, a robust feature across multiple observing runs is that detected BHs exhibit predominantly low spins ($a_*\sim 0.0$--$0.2$) \citep{Abbott23}. This conclusion persists in the most recent GWTC--4.0 results \citep{LVK25b} and is consistent with several independent re-analyses of the current GW population \citep[e.g.,][]{Adamcewicz2025}.

Furthermore, \citet{Dhani25} show that waveform inaccuracies can artificially inflate the inferred spin magnitudes, producing posteriors that favor high spins even when the true spins are modest. These biases are most pronounced for binaries with unequal masses and strong precession, implying that spin magnitudes may be systematically overestimated in precisely the systems where high spins would be most informative. Consequently, the already low incidence of high spins in the GW population may be even sparser once waveform-modeling systematics are fully accounted for.

\section{Results from X-ray binaries}
\label{EM}

X-ray spectra from XRBs consist of three main components (see \citealt{DGK07} for an exhaustive review). The first is thermal, blackbody-like emission from an optically-thick and geometrically thin accretion disk \citep{SS73}. The characteristic maximum temperature of the disks is $\sim$1 keV, which means they cannot account for the emission commonly seen in hard X-rays. That emission is instead well explained by the second component: Compton upscattering of the soft photons from the disk blackbody, possibly including an additional source of soft seed photons from synchrotron emission (e.g., \citealt{WZ00, VVP11}). Then, the third component is due to the Comptonization emission being reflected from the disk, which forms characteristic spectra due to Compton backscattering and atomic absorption/reemission \citep{LW88, Fabian89, GK10}. A prominent feature of those spectra is the Fe \textrm{K}$\alpha$ fluorescence line \citep{Fabian89}. 

BH XRBs have two main luminous spectral states \citep{DGK07}. In the soft state, the disk emission dominates, and the high-energy tail is relatively weak. In the hard state, the disk emission is weak, and the dominant emission is from Comptonization. 

All three aforementioned components are relativistically modified due to the Doppler and General Relativistic effects \citep{Fabian89, Dauser13}. The thermal disk spectra are also modified by radiative transfer in their upper layers \citep{Davis05}, and the disk temperature increases with $a_*$. Moreover, the spin value also affects the timing and polarization properties of the observed emission. Finally, the power of the jets launched by the accretion flow and the BH depends on the spin. We discuss these effects in detail below. 

\subsection{Spectral methods of spin measurements}
\label{spectral}

\subsubsection{The reflection method}
\label{reflection}

One of two main methods used for the spin determination of BH XRBs (with either high and low-mass donors) consists of fitting their X-ray spectra taking into account a component due to the reflection from the disk, including the X-ray Fe K$\alpha$ fluorescent line at the rest energy of 6.4--7.0\,keV (see \citealt{Bambi21} for a review). This {\it reflection method\/} utilizes the effects of gravitational redshift and Doppler shift acting on the rest-frame spectrum, and it relies on the strong dependence of these effects on the disk emitting radii, which in turn depend on $a_*$. Its important advantage is in the independence of the BH mass and the distance. However, to be effectively used, the disk has to extend to the innermost stable circular orbit (ISCO), or at least be close to it. This is because the space-time metric at radii $\gg\! R_{\rm g}$ (where $R_{\rm g}\equiv G M/c^2$ is the gravitational radius) becomes almost independent of $a_*$. Thus, this method could, in principle, be reliably used in the soft spectral states of BH XRBs (where the disk most likely extends to the ISCO and where reflection is more pronounced; \citealt{Steiner16}) as long as their high-energy spectral tails are sufficiently strong.

\begin{figure}
\centerline{\includegraphics[width=8cm]{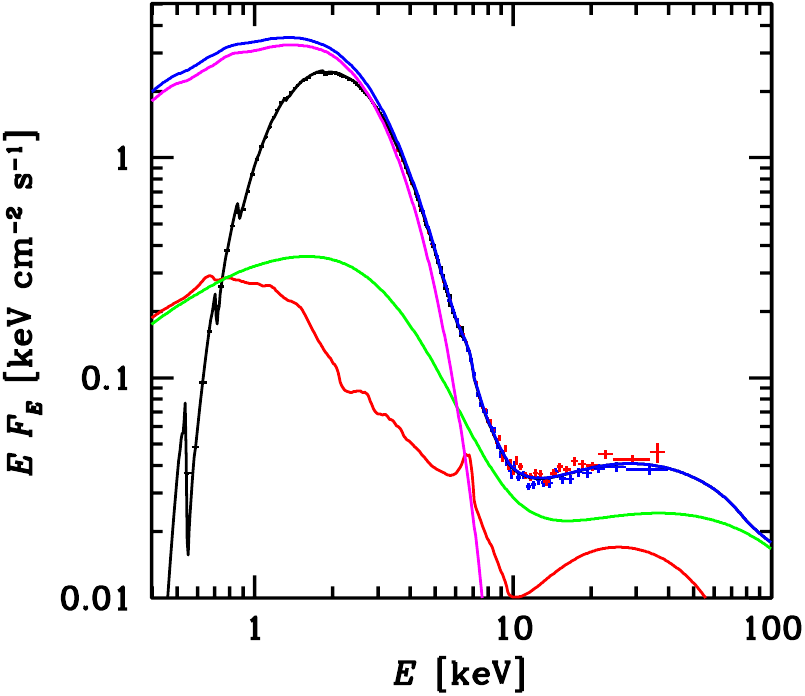}} 
\caption{The NICER (black crosses) and NuSTAR A and B (blue and red crosses, respectively) spectra of GX 339--4 fitted by the relativistic disk model {\tt slimbh} (magenta), Compton upscattering of the disk emission (green) and its reflection from the disk (red), with the total spectrum shown by the blue curve. The spectrum below a few keV is attenuated by the Galactic absorption, as shown by the black curve. We see that the Comptonization spectrum (green), which is both emitted toward the observer and incident on the disk (thus giving rise to the reflection component), is strongly curved and very different from a power law. Adapted from \citet{Zdziarski25a}.
}\label{gx339}
\end{figure}

The origin of the high-energy tail is most likely Compton upscattering (Comptonization) of the disk emission by a population of coronal high-energy electrons (e.g., \citealt{G99}). However, as pointed out in \citet{Zdziarski25a}, the soft-state Comptonization spectra, emitted both outside and toward the disk, do not usually have the form of a power law with a high-energy cutoff (as assumed in most of the relativistic reflection models). This is illustrated in Fig.\ \ref{gx339}, showing a soft-state X-ray spectrum of the XRB GX 339--4 and its model components \citep{Zdziarski25a}. We see that the Comptonization spectral component is strongly curved. Its curvature can then strongly affect the fit results using the power-law approximation; in particular, the Fe K$\alpha$ line around 6--7 keV is much stronger than it would be if the incident spectrum were a power law fitted to the high-energy tail at $\gtrsim$ 10 keV. Thus, the reflection method in the soft state should be used with the convolution over the incident spectra, as it was done in the example in Fig.\ \ref{gx339}. 

There are a few convolution codes available within the X-ray fitting suite {\sc xspec} \citep{Arnaud96}, in particular {\tt xilconv} \citep{Kolehmainen11}, which is based on the opacity tables of the rest-frame reflection code {\tt xilver} \citep{Garcia13} combined with the Green's functions of \citet{MZ95}. An alternative is {\tt rfxconv}, which uses instead the opacity tables of the reflection code {\tt reflionx} \citep{Ross99, Ross07}. While the use of the convolution method with curved spectra is their major advantage, allowing us to estimate the reflection spectra from an arbitrary incident spectrum, their accuracy remains limited, see \citet{Ding24}. Furthermore, they are one-zone models, and the current version of {\tt xilconv} uses the tables of the version of {\tt xilver} of \citet{Garcia13} (before v.1.0) in which the electron density is fixed at $10^{15}$ cm$^{-3}$. The resulting rest-frame reflection spectra have to be then relativistically broadened, e.g., using the code {\tt relconv} \citep{Dauser10}.

On the other hand, the reflection method of spin measurement has been applied to sources in the hard spectral state (e.g., \citealt{Garcia18, Draghis24}), which is controversial because the extension of the disk to the ISCO in that state is highly uncertain (e.g., \citealt{Basak16, Mahmoud19, Kawamura22, Kawamura23}). This is because the shape of the Fe K$\alpha$ feature is obtained by subtracting the shape of the primary continuum, and thus, it critically depends on the latter. Often, it is assumed to be a single component, e.g., a power law or a thermal Comptonization spectrum. However, there is strong evidence for the complexity of the primary spectra in the hard state, with at least two dominant Comptonization components, harder and softer (e.g., \citealt{Basak17, Zdziarski21b, Zdziarski21c, Kawamura22, Chand24, Sahu25}). In that case, the presumed red wing of the Fe K$\alpha$ line becomes a part of the softer incident continuum, which, in turn, results in the spectral fits yielding truncated disks.  

Another caveat for many studies using this method is the adopted assumption that the irradiating source forms a so-called lamppost, a point source on the BH spin axis at some distance from the BH \citep{Martocchia96, Miniutti04}. In contrast, recent X-ray polarization data suggest the hot irradiating plasma is spatially extended perpendicular to the disk axis, consistent with either a slab coronal geometry (which would be the case, in particular, in the soft state) or the presence of a hot inner flow, see, e.g., \citet{Krawczynski22} for Cyg X-1, and \citet{Veledina23}, \citet{Ingram24}, \citet{Podgorny24} for Swift J1727.8--1613. There are many challenges to models where polarization signatures are produced above the BH (e.g., in a current sheath; \citealt{DexterBegelman2024, Sridhar2025}). Namely, the high and stable polarization degree (PD), the stability of the polarization angle (PA) across different states and luminosities of the same source, its persistent alignment with the jet direction, or even the similarity of polarization signatures across different sources of the same inclination \citep{Kravtsov2025}. Moreover, these models are highly sensitive to system parameters such as inclination, optical depth of matter, and BH spin. Existing constraints coming from polarization data thus favor a simple, slab-like or wedge-like geometry of the Comptonization region, with PD increasing with inclination, which is indeed observed \citep{Ewing2025}.

Another issue affecting the accuracy of the reflection method is the dependence of the reflection spectra on the irradiating flux, and through that, on the density of the reflector, since a fitting parameter of the reflection codes is the ionization parameter
\begin{equation}
\xi\equiv 4{\pi}F_{\rm{irr}}/n,
\label{xi}
\end{equation}
where $F_{\rm {irr}}$ is the flux irradiating on the reflector and $n$ is the electron density. Thus, the higher the flux, the higher the density at a given fitted value of $\xi$. Many studies have been done assuming the density of $10^{15}$ cm$^{-3}$ \citep{Garcia13}, which is applicable to AGN studies. The current version of {\tt xillver} works for $n$ from $10^{15}$ to $10^{20}$ cm$^{-3}$, but the disk densities in BH XRBs can be higher. There is an ongoing effort to improve the accuracy of {\tt xillver} further \citep{Ding24}. As \citet{Ding24} note, a statistically excellent fit does not guarantee the correctness of the model parameters, in particular because there are many local minima in a multi-dimensional spectral fitting procedure. Another important caveat is that the reflecting slab in those codes is assumed to be heated only by the irradiating flux. However, disks in the soft state are strongly viscously heated, with that heating dominating the effect of the irradiation.

Another potentially important effect for measuring spins is the reflection of blackbody radiation returning to the disk owing to the BH gravity. This effect is included in some way in {\tt kerrbb} and {\tt kerrbb2} (see below), which allow the user to switch on the self-irradiation, and it has also been included in recent reflection modeling. However, it is assumed that the disk is completely absorbing (i.e., with null albedo), which then only slightly increases the local temperatures. \citet{Schnittman09} present calculations of the spectra and polarization of that component, assuming the disk is completely reflecting (i.e., with albedo $=1$). It then results in a weak and soft high-energy tail beyond the disk blackbody, with its amplitude and polarization degree increasing with the spin. The tail becomes noticeable only at $a_*\gtrsim 0.9$ \citep{Schnittman09}. Then, \citet{Taverna21} presented calculations of that effect as a function of the disk ionization. On the other hand, \citet{Connors20, Connors21} included an irradiating blackbody component with free normalization and temperature in their reflection fits (i.e., without including their dependencies on the spin of \citealt{Schnittman09}), and attributed it to the returning disk blackbody. Finally, \citet{Dauser22} and \citet{Mirzaev24} performed calculations of the returning coronal reflection.

\subsubsection{The continuum method: standard models}
\label{continuum}

The other main method used for spin determination of BH XRBs is the {\it continuum method\/} \citep{McClintock14}, which is based on fitting the X-ray shape of the disk continuum. The method assumes the disk extends down to the ISCO, which appears to be well-established observationally in the soft spectral state (e.g., \citealt{GD04, Steiner11}). Also, it assumes the applicability of the standard disk models \citep{SS73, NT73, Abramowicz88, Sadowski09}. This method, in principle, requires knowledge of the BH mass and the distance to the source. It relies on the fast decrease of the ISCO radius with the increasing BH spin \citep{Bardeen72}. High-energy tails, commonly appearing beyond the disk spectra, are attributed to the Comptonization of the disk photons in a corona above the disk and the reflection from the disk. Then, the continuum and reflection methods can be used together for the same data set, as done by \citet{Miller09}, \citet{Parker16}, and \citet{Zdziarski24b, Zdziarski25a}. 

\renewcommand{\arraystretch}{1.15}
\begin{table*}
\caption{Relativistic disk models}
\vskip 0.2cm
\centering\begin{tabular}{lccccccc}
\hline
Model & {\tt kerrbb} & {\tt kerrbb2} & {\tt bhspec} & {\tt bhspec} & {\tt slimbh} & {\tt slimbh} & {\tt kynbb} \\
Method &free $f_{\rm col}$ & fitted $f_{\rm col}$ &atmosphere  & atmosphere & free $f_{\rm col}$ & atmosphere &free $f_{\rm col}$\\
\hline
$\alpha$ & -- & 0.01--0.1 & 0.01, 0.1 & 0.01 &  -- & 0.005--0.1 & -- \\
$a_*$ & $-1$--0.9999 & $-1$--0.99 & 0--0.8 & 0--0.99  & 0--0.999 & 0--0.999 & $-1$--1 \\
$L_{\rm d}/L_{\rm Edd}$ & $\lesssim$0.3 & 0.03--0.3 & 0.01--1  & 0.01--1 & 0.05--1 & 0.05--1 & $\lesssim$0.3\\
\hline
\end{tabular}\\
\label{disk_models}
\vskip 0.2cm
{{\it Notes:} A version of {\tt bhspec} covering $a_*$ of $-1$--0.99 is available from Shane Davis upon request. The normalization should be kept at unity in all of the models except in {\tt bhspec}, where it is tied to the source distance. Note that {\tt kerrbb} is agnostic of the value of $L_{\rm d}/L_{\rm Edd}$ implied by its parameters, and thus the condition $L_{\rm d}/L_{\rm Edd}\lesssim 0.3$ needs to be checked a posteriori. The $f_{\rm col}$ of {\tt kerrbb2} is obtained by fitting its values in {\tt kerrbb} to the corresponding spectra generated from {\tt bhspec}. }
\end{table*}

A significant issue regarding the continuum method is the local rest-frame spectra of the disk. Table \ref{disk_models} lists the methods used by the main available models and the ranges of the main parameters, $a_*$, the viscosity parameter, $\alpha$, and the Eddington ratio, $L_{\rm d}/L_{\rm Edd}$, where $L_{\rm d}$ is the disk luminosity, and $L_{\rm Edd}$ is the Eddington luminosity for hydrogen. A commonly used model {\tt kerrbb} \citep{Li05} approximates the local spectra as blackbodies with a color correction, $f_{\rm col}$ \citep{ST95, Davis19}, where this factor is defined as the ratio of the color temperature to the effective one, $f_{\rm col}\equiv T_{\rm color}/T_{\rm eff}$. The value of $f_{\rm col}$ for a given disk is uncertain, though usually in the range of 1.5--2 \citep{Davis19}. The same parametrization is used by the model {\tt kynbb}\footnote{\url{https://projects.asu.cas.cz/stronggravity/kyn}} \citep{Dovciak04b}, which also allows the inner disk radius to be different from the ISCO radius, $R_{\rm ISCO}$. 

However, disk spectra from the standard thin disk \citep{NT73} are significantly more complex than diluted blackbodies when taking into account radiative transfer (e.g., \citealt{Davis05}). The models {\tt bhspec} and {\tt slimbh} use the atmospheric calculations of \citet{Davis05} and \citet{Davis06} (which use the viscosity parametrized by $\alpha$; \citealt{SS73}) to describe the rest-frame spectra, with the latter also taking into account the finite disk thickness \citep{Sadowski09, Sadowski11a, Sadowski11b, Straub11}. Thus, {\tt slimbh} appears to be the most advanced relativistic disk model available for spectral fitting. Its main limitations are that it does not cover negative spins and low Eddington ratios (see Table \ref{disk_models}). Still, the radiative transfer calculations of \citet{Davis05} and \citet{Davis06} were done using the standard disk model, assuming its stability, while the disk is unstable in most of the regimes calculated by them.

Then, the model {\tt kerrbb2}\footnote{\url{https://jfsteiner.com/?p=55}} \citep{McClintock06} is a version of {\tt kerrbb} in which the values of $f_{\rm col}$ have been fitted to the spectra of \citet{Davis05, Davis06}. Note that {\tt kerrbb} and {\tt kerrbb2} use as a parameter the disk mass accretion rate, $\dot M_{\rm d}$, instead of $L_{\rm d}/L_{\rm Edd}$. They are related by
\begin{equation}
\frac{L_{\rm d}}{L_{\rm Edd}}=\frac{\eta(a_*)\dot M_{\rm d} c \sigma_{\rm T}}{4\pi G M m_{\rm p}},
\label{eddratio}
\end{equation}
where $\sigma_{\rm T}$ is the Thomson cross section, $m_{\rm p}$ is the proton mass, $\eta(a_*)$ is the accretion efficiency \citep{Cunningham75},
\begin{equation}
\eta(a_*)=1-\left[1-\frac{2}{3 r_{\rm ISCO}(a_*)}\right]^{1/2}, 
\label{efficiency}
\end{equation}
and $r_{\rm ISCO}(a_*)\equiv R_{\rm ISCO}(a_*)c^2/GM$, see equation (2.21) of \citet{Bardeen72}\footnote{Note that while Equation (\ref{efficiency}) is significantly simpler than the formula for the efficiency based on Equation (2.12) of \citet{Bardeen72}, they are equivalent.}. We note that, since {\tt kerrbb} assumes the thin-disk model, its fit results are independent of the Eddington ratio. Therefore, the fitted spectra remain unchanged with varying $M$, $D$, and $\dot M_{\rm d}$ as long as the scaling  $M\propto  D\propto \dot M_{\rm d}^2$ is preserved.

Finally, we notice that the simple nonrelativistic model of the disk continuum, {\tt diskbb} \citep{Mitsuda84}, is often used to roughly estimate the spin. The model gives an approximate (and proportional to the distance) estimate of the disk's inner radius, which can be converted to the spin if the mass and distance are known. This model follows the formulation of \citet{SS73} {\it except\/} that it does not include the zero-stress boundary condition at the ISCO, $(1-\sqrt{R_{\rm ISCO}/R})$. The distance, $D$, inclination, $i$, and the inner radius are related to the {\tt diskbb} normalization, $N_{\rm dbb}$, by
\begin{equation}
R_{\rm in}=10^5 f_{\rm in} f_{\rm col}^2 \frac{D}{10\,{\rm kpc}} \left(\frac{N_{\tt dbb}}{\cos i}\right)^{1/2}\,{\rm cm},
\label{rin_disk}
\end{equation}
where $f_{\rm in}<1$ is a correction factor accounting for the lack of the zero-stress term at the ISCO. Assuming $R_{\rm in}=R_{\rm ISCO}$, we can calculate the spin for known $D$ and $M_1$ using the formula of \citet{Bardeen72}. The lack of knowledge of the precise values of $f_{\rm in}$ is a major drawback of this method. While \citet{Kubota98} estimated $f_{\rm in}\approx 0.41$, this is valid only for $a_*=0$, while the method is commonly used to determine $a_*\neq 0$. Also, their estimate is nonrelativistic, making the estimates of $a_*$ quite imprecise. Still, {\tt diskbb} can be used to compare the inner radii of the same source in different flux states, to determine whether the inner radius remains constant or not. 

\subsubsection{The continuum method: problems and alternatives}
\label{problems}

It is well known that the standard disk model fails to describe some important astrophysical phenomena. Most importantly for us, it predicts the disk to be viscously and thermally unstable when dominated by radiation pressure (\citealt{Lightman74, Shakura76}, see also \citealt{Blaes2025}). The instabilities cause the disk to break up into rings as well as undergo runaway heating/cooling. On the contrary, observations of the X-ray emission in the soft states of BH XRBs show their disks to be very stable \citep{GD04} for $L\lesssim L_{\rm Edd}$, i.e., including the radiation-pressure dominated regime. Thus, the standard disk model, while predicting spectra consistent with observations, is {\it not\/} a self-consistent and correct description of the disks in the soft states of BH XRBs. 

Moreover, the standard model predicts disk gravitational fragmentation in the case of AGNs, which is not observed (e.g., \citealt{Begelman07}). We also note that the disk sizes in AGNs inferred from microlensing are larger by a factor of a few than those predicted by the standard disk theory, as seen in \citet{Chartas16}.

The inclusion of globally organized magnetic fields in the structure of the accretion flow proved to be a significant development in this field. This has been done either by considering the additional torques imposed by the resulting \citet{BP82} jets \citep[see, e.g.,][and references therein]{Ferreira95, Ferreira97}, or more simply by the inclusion of the magnetic pressure to stabilize $\alpha$-disks \citep[see, e.g.,][]{Begelman07}. The impact of the global magnetic field has since been confirmed in numerous numerical simulations \citep[see, e.g.,][]{Sadowski16a}, even in the case of weakly magnetized disks \citep[e.g.,][]{Jacquemin-Ide19}. It was also found that a necessary ingredient for that was the presence of a poloidal field, which led to an amplification of the toroidal field \citep[][and references therein]{Salvesen16b, Sadowski16b, Jacquemin-Ide24}. The stability was confirmed by simulations for a wide range of magnetic field configurations \citep{Mishra22}.

An example of this class of models is the so-called magnetically elevated disks \citep{Begelman07, Begelman17, Mishra20}. Importantly, they have lower gas densities at the effective photosphere, which enhances electron scattering and could lead to a harder spectrum than from a disk without magnetic pressure support. For this reason, their average color correction could be significantly larger than the standard range of $f_{\rm col}\approx 1.5$--$2$. No radiative transfer calculations of the spectra from magnetically elevated disks have, to our knowledge, been done yet. Still, there have been recent advancements in numerical simulations of accretion flows, including more realistic modeling of radiation transfer \citep[e.g.,][]{Sadowski_Narayan16, Huang23, Fragile23b} as well as more sophisticated tools for generating spectral predictions \citep[e.g.,][]{Kinch19, Mills24, Roth25}.

\begin{figure}[t!]
\centerline{\includegraphics[width=1.03\columnwidth]{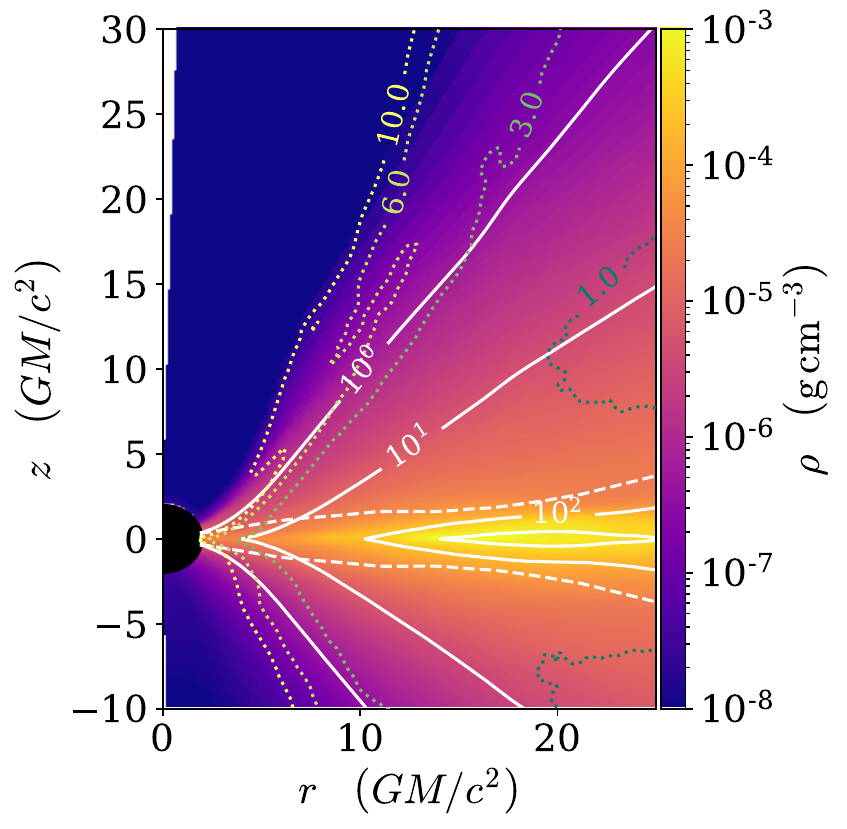}}
    \caption{The vertical structure of a puffy disk around a non-rotating BH at $\dot M_{\rm accr}\approx 0.6 \dot M_{\rm Edd}$ averaged over time and azimuth. The color map shows the density, and the dashed white curve gives the disk scale height, defined by $\left[\int \rho(r,z) z^2{\rm d}z/\int\rho(r,z){\rm d}z\right]^{1/2}$. The dotted green-to-yellow curves give the contours of the gas temperature, $kT_{\rm e}$ in keV. The solid white curves show the $z$-integrated optical depth for $\tau_{\rm T} = 1$ (the photosphere), 10, 100, and 300. We see the presence of a surface layer with $\tau_{\rm T}\sim 10$ at the temperature of $kT_{\rm e}\sim 1$ keV, which can be interpreted as a warm corona. We note that the flow does not experience any noticeable change at the ISCO radius, $6\, G M/c^2$. Based on \citet{Lancova19}.}
\label{puffy_structure}
\end{figure}

Alternatively, a significant part of the total pressure support could be provided by a radial magnetic field, in addition to the toroidal one \citep{Lancova19}, resulting in a thicker disk, with a photosphere that is hotter than that of a standard disk, see figure 2 in \citet{Wielgus22}. Radiative general relativistic magnetohydrodynamics (GRMHD) simulations of these so-called puffy disk models have been considered in the regime $\dot M_{\rm accr} \approx \dot M_{\rm Edd}$ (where we define $\dot M_{\rm Edd}\equiv L_{\rm Edd}/0.1 c^2$). An example of the vertical structure of a puffy disk is shown in Fig.\ \ref{puffy_structure}, where the photosphere is located at around $z \approx r$ (solid white line at $\tau_{\rm T} =1$), while the disk scale height is $z \approx 0.1 \, r$ (dashed white line). The presence of a hotter and geometrically thick photosphere reaching to the BH event horizon results in a significant overestimate of the BH spin. When the {\tt kerrbb} and {\tt slimbh} models were used on synthetic spectra obtained from the simulations assuming $a_*=0$, the fitted spin values ranged from $a_* \approx 0.5$ to $\approx$0.9 \citep{Wielgus22, Lancova23}. While these numerical simulations, like all others, come with numerous caveats, this case study is illuminating because it highlights how the more complex physical behavior seen in simulations (and potentially occurring in nature) can lead to significant biases in BH spin estimates, even when interpreted with state-of-the-art spectral models.

In a related development, warm coronae (with the electron temperature of $kT_{\rm e}\sim 1$\,keV and the Thomson optical depth of $\tau_{\rm T}\gtrsim 10$) on top of optically thick accretion disks have been invoked to explain soft X-ray excesses found in many AGNs (e.g., \citealt{MBZ98, Petrucci20, Ursini20, Xiang22}, see however \citealt{KaraGarcia25}). Warm coronae have also been found to be necessary to explain the broadband spectra of AGNs in their soft and intermediate states (analogous to those of BH XRBs), see \citet{Hagen24, Kang25}. A theoretical model for warm coronae above XRB disks formed by magnetic heating was developed by \citet{Gronkiewicz20}. A warm coronal region with $\tau_{\rm T}\sim 10$ and $kT_{\rm e}\sim 1$ keV is seen in the simulation shown in Fig.\ \ref{puffy_structure}. 

What is the origin of the Fe K$\alpha$ line in this model? The coronal temperature is low enough for the ionization of the corona to be incomplete, and Fe partial ions can absorb the incident radiation, which is followed by fluorescence, as shown by \citet{Biswas25}. Additionally, the corona is likely to cover the disk only partially, as is the case in AGNs \citep{Kang25}. Then, some reflection and fluorescence occur from the uncovered disk.

Typically, the fitted coronal temperature is approximately the same as the inner disk temperature. This prevents the formation of a power-law spectral component due to Comptonization of the disk photons by those electrons. Instead, the overall spectral shape is modified. Together with the Galactic absorption being strong at $\lesssim$\,1 keV, these effects complicate placing effective spectral constraints on the existence of warm coronae in XRBs. Nevertheless, its presence was hinted at in some BH X-ray binaries (BH XRBs), particularly in the soft state of GRS 1915+105 (a BH X-ray binary with a low-mass donor, hereafter LMXB), which was fitted by a hybrid Comptonization model including a low-temperature, high optical depth thermal component \citep{ZGP01}. A similar result was found for the very high state of GRO J1655--40 \citep{Kubota01}. 

Following those results, models with warm coronae were fitted by \citet{Belczynski24} and \citet{Zdziarski24a, Zdziarski24b, Zdziarski25a} to the soft states of LMC X-1, Cyg X-1, M33 X-7, and GX 339--4. As we discuss in Section \ref{revised} below, this model provides statistically good fits and significantly alters the fitted values of the spin. However, the one-zone Comptonization model used so far for the warm corona \citep{Z20_thcomp} does not account for the detailed corona structure presented in \citet{Gronkiewicz20}. 

Another promising development involves considering the contribution from the continuum emission of the plunging region, i.e., the region below the ISCO. This component has generally been neglected in standard spectral models, including all those discussed in Section \ref{continuum}, which is justified for standard disk models based on \citet{NT73}, for which the stress at the ISCO is close to zero. However, if the angular momentum transport in the disk is driven by the magnetorotational instability \citep[MRI, ][]{Balbus91}, which is now generally accepted, even a small-scale magnetic field embedded in the disk will result in significant Maxwell stress on the ISCO. The possible presence of such stress was first mentioned by \citet{Thorne74}. \citet{Reynolds97} was the first to consider the effect of the intra-ISCO emission on the profile of the fluorescence Fe K$\alpha$ line, which was later studied by \citet{Schnittman13}. Quantitative treatments of such effects were done, e.g., by \citet{Hawley02}, \citet{Krolik05}, and \citet{Noble10, Noble11}. Calculations of the effect of the intra-ISCO emission were done, e.g., by \citet{Zhu12} and \citet{Mummery25b}.

Then, both analytical work and numerical simulations show that the viscous $\alpha$ parameter cannot be taken as constant, typically rising steeply near the ISCO and reaching a sharp maximum within the plunging region \citep{Krolik99, Agol00, Sorathia12, Penna13, Abramowicz96, Jiang19}. Thus, this increase of stress will also affect the flow within the plunging region and reduce the importance of the ISCO, simply as a consequence of the MHD properties of the accreting fluid. Numerical simulations confirmed that the ISCO plays no important role in flow dynamics (see Fig.\ \ref{puffy_structure}, figure 3 in \citealt{Rule25}, and \citealt{LancovaInPrep}). The disk inner edge, as the location up to where the disk radiates as a blackbody, should be the (magneto)sonic point, rather than the ISCO \citep{Sadowski09}. For low mass accretion rates, these coincide \citep{Armitage01, Penna12}, but deviate strongly with increasing mass accretion rate. Results of simulations showed that the photosphere location and the density scale height may also differ significantly, in contrast to the assumption of analytical disk models, and even for low luminosities \citep[e.g., ][]{Curd23, Liska22, Mishra22}.

The non-zero stress at ISCO may be parametrized by $\delta_{\rm j}$, defined as the ratio of the angular momentum passed back to the disk at ISCO to the angular momentum at ISCO, which approximately equals $\alpha (H/R)$ \citep{Paczynski00, Mummery23}. However, there is a tension between what value it may reach (ranging from a fraction of a percent to several percent) and how much it may affect the spectra produced within the plunging region (compare results of \citealt{Mummery25b} to those of \citealt{Zhu12}). Numerical simulations yield different values, which may be caused by the differences in the simulated regime or the physics included. Those of \citet{Zhu12} can be fitted with $\delta_{\rm j}\sim 0.5\%$, while \citet{Rule25} report value of $\delta_{\rm j}\sim 5\%$, and the puffy disk simulations (which includes radiation and target higher mass accretion rates) give $\delta_{\rm j}\sim 16\%$ \citep{LancovaInPrep}. Furthermore, \citet{Lasota24} argued that this value cannot be taken as a free parameter, since the solution would lead to an unphysical velocity on the BH horizon.

Using an extended trans-ISCO thin disk model, \citet{Mummery25b} showed that the emission from the plunging region of a Schwarzschild BH mimics the effect of a high spin. This was also demonstrated by detailed spectral fitting of the X-ray spectra of the HMXB M33 X-7 with {\tt fullkerr}, a new model developed by \citet{Mummery24}\footnote{\url{ https://github.com/andymummeryastro/fullkerr}}, assuming $a_*=0$, obtaining $\delta_{\rm J}=0.078$. This fit was slightly better than that using {\tt kerrbb} (i.e., at $\delta_{\rm J}=0$), which yielded $a_*=0.84\pm 0.05$. We note that \citet{Mummery25b} modeled the emission from $R>R_{\rm ISCO}$ by the standard model \citep{NT73} except for the effect of the finite stress at the ISCO taken into account following \citet{Agol00}, but the issue of the thin disk model instabilities in a radiation-pressure-dominated regime is not discussed within their work.

An associated effect is the presence of a limit on spin due to accretion. When magnetic fields are present, the stress continues to the horizon. This results in a strongly spin-dependent outward flux of angular momentum conveyed electromagnetically. \citet{Gammie04} and \citet{Krolik05} found that this effect limits the spin to $a_*\lesssim 0.9$. A weaker effect is the capture of photons emitted within the ISCO. The presence of magnetic fields modifies the standard limit of $a_{*{\rm max}}= 0.998$ of \citet{Thorne74} to $a_{*{\rm max}}\lesssim 0.99$ for typical values of the stress at the ISCO found in numerical simulations \citep{Mummery25a}. 

An important question in the above models is whether they satisfy the $L\propto T_{\rm max}^4$ correlation found in many sources in the soft state \citep{GD04}, where $T_{\rm max}$ is the maximum fitted disk temperature. This correlation suggests a constancy of the disk's inner radius with changing luminosity, pointing to the importance of the ISCO. This would then be an important test of the validity of any modified disk model.

Moreover, all of the disk-continuum models discussed above assume that the normal to the disk surrounding the BH and emitting the fitted blackbody radiation is aligned with the BH spin. However, the BH spin in some XRBs can be misaligned with the binary axis, e.g., in MAXI J1820+070 \citep{Poutanen22} or Cyg X-3 \citep{Dmytriiev24}. Also, the currently leading model of low-frequency quasi-periodic oscillations (hereafter QPOs) in the hard state \citep{Fragile07, Ingram09} predicts that all XRBs with low-frequency QPOs would have a misalignment between the binary and BH axes. 

On the other hand, the inner disk can remain aligned with the BH spin even in the presence of a BH spin-binary axis misalignment \citep{Bardeen75}. Such geometry, which is expected in case of a spin-orbit misalignment in X-ray binaries \citep{Nixon12, Marcel21}, appears, in several cases, to be supported by measurements of the X-ray polarization. In all cases in which there are both such measurements and resolved radio jets, the polarization angle is aligned with the jet position angle (e.g., Cyg X-1, \citealt{Krawczynski22}; MAXI J1727.8--1613, \citealt{Wood24}) or perpendicular (Cyg X-3, \citealt{Veledina24, Veledina24b}). Assuming that the jets are produced by the spin-extraction mechanism (\citealt{BZ77}; see Section \ref{jet}), these results are compatible with the inner disk normal being aligned with the BH spin direction. 

However, the fitted inclination could be different from the binary inclination in some cases. Then, a misalignment between the inner disk and the BH spin would significantly affect the fitted results. The unknown, but expected, misalignment between the binary axis and the BH spin implies yet another caveat to the disk-continuum fitting methods as well as reflection studies.

\subsection{QPOs and spins}
\label{QPO}

Yet another method derives the spin from QPOs. QPOs are narrow features observed in the X-ray power spectra of accreting compact objects. They are thought to reflect coherent, albeit transient, processes in the inner accretion flow. QPOs are typically categorized by frequency into low-frequency ($\lesssim 30$ Hz) and high-frequency ($\gtrsim 60$ Hz) types. Multiple models have been proposed to explain the most commonly observed ones at low frequencies, ranging from geometric precession of the inner flow to disk oscillation modes and relativistic test-particle dynamics; see \citet{Ingram19} for a review. Among these, the Relativistic Precession Model (RPM) provides a framework for connecting all observed QPOs (both high and low-frequency) with the fundamental frequencies of particle motion around a spinning BH, offering a potential method to infer the BH spin and mass.

Current versions of this method are based on several theoretical models of disk oscillations occurring at special radii of the disk and rely on pure test-particle dynamics. The existing models identify high-frequency QPOs with either a resonance between the orbital and radial epicyclic frequencies \citep{Abramowicz01} or some other local disk oscillations. In particular, the QPO frequencies are identified with those of the disk oscillations by the RPM \citep{Stella98, Stella99, Stella99b}. It offers a conceptually straightforward method to estimate BH spin by associating observed QPOs with fundamental frequencies of test-particle motion in the accretion flow: the periastron precession ($\nu_{\rm per}$), the orbital frequency ($\nu_{\phi}$), and the Lense-Thirring \citep{LT18} frequency, $\nu_{\rm LT}$, sometimes called the nodal frequency, $\nu_{\rm nod}$. See \citet{Motta14} and \citet{Ingram19} for detailed reviews. It is crucial to distinguish the RPM model from the solid-body Lense-Thirring precession models that consider the global precession of the inner flow to explain the low-frequency QPOs (\citealt{Fragile07, Ingram09}; see discussion in \citealt{Motta18}). Unlike solid-body precession models, the RPM does not rely on spin-orbit misalignment but instead on small perturbations to an otherwise circular, aligned disk. The key limitation of this model is that any perturbation of a small disk element (``test particle'') would be rapidly damped by interactions with the surrounding material, preventing the establishment of coherent relativistic precession. The RPM frequencies depend solely on the BH mass, spin, and the radius at which the oscillation originates -- typically assumed to be the transition/truncation radius between the cold accretion disk and the hot accretion flow. In the innermost regions of a stellar-mass BH accretion flow, these frequencies lie in the $\sim 0.1$--$10^3$ Hz range.

In rare cases where three (two high-frequency, one low-frequency) unrelated QPOs are observed simultaneously, the system of equations in the RPM can be solved analytically for mass, spin, and radius. Such triplets have been reported at least twice, each time yielding low spin values: $a_* = 0.290 \pm 0.003$ for GRO J1655--40 \citep{Motta14} and $a_* = 0.149 \pm 0.005$ for XTE J1859+226 \citep{Motta22}, with the inferred radii of $R = 5.65 R_{\rm g}$ and $6.85 R_{\rm g}$, respectively, and the BH masses consistent with the dynamical estimates. However, this method suffers from two major limitations: (1) the physical interpretation of the derived radius is unclear, and its association with the truncation radius is speculative; and (2) QPO triplets are exceedingly rare around stellar-mass BHs. To address these issues, \citet{Franchini17} proposed focusing on the soft states, where the inner disk is likely at the ISCO and where the difference between the local Lense-Thirring frequency and that of a precessing hot flow is negligible due to the small size of the hot flow, thus providing some physical justification. Assuming that observed low-frequency QPOs in such states originate at the ISCO and correspond to $\nu_{\rm LT}$, they constrained the spin values for a dozen sources (assuming $M = 3$--$20 \msun$), finding all constrained values to lie in the range of $a_* \approx 0.08$--$0.47$. However, those QPOs are only observed in the hard and intermediate states of XRBs, and, to our knowledge, no convincing justification for the absence of QPOs in the soft states has been provided in this framework.

While this method is appealing both by its simplicity and its consistency with the spins obtained from GW signals, we must remember that there is no yet physical justification for the process producing the associated QPOs (unless one lies close to the ISCO, where, however, dissipation is close to null). Furthermore, it requires the presence of slightly elliptical orbits (so that $\nu_{\rm per} > 0$), which remains to be demonstrated, and non-zero BH spins (so that $\nu_{\rm LT} > 0$). Moreover, in the case of neutron stars, where high-frequency QPOs are more common, the relationship between test-particle frequencies and the QPOs remains unclear (see, e.g., \citealt{Mendez07, Doesburgh17}).

\subsection{Jet power and spins}
\label{jet}

The spin can also be approximately constrained from the jet power in the model of BH spin extraction \citep{BZ77}. The highest possible jet power found from GRMHD simulations for a given $a_*$ is (\citealt{Davis20} and references therein),
\begin{equation}
P_{\rm j}\approx 1.3 \dot M_{\rm accr}c^2 a_*^2,
\label{Pj}
\end{equation} 
where $\dot M_{\rm accr}$ is the total (for both the jet and counterjet) accretion rate. This limit follows from the magnetic pressure balanced by the accretion ram pressure, which case is called the Magnetically Arrested Disk (MAD; \citealt{BK74, Narayan03, McKinney12}). Thus, $a_*\sim 1$ is needed to achieve the maximum jet power. 

Transient ejecta achieve the highest power among jets launched by BH XRBs during hard-to-soft state transitions, with a prominent example of MAXI J1348--630 \citep{Carotenuto21}. The jet power has been estimated based on the kinetic energy obtained from modeling the propagation of those jets (observed in radio) up to distances up to $\sim$1 pc through the surrounding media, together with the jet launching times assumed to equal the durations of major radio flares accompanying the ejecta. Typically, the major flares are usually well time-resolved, and last $\lesssim$1 day \citep{Carotenuto21, Cooper25}. The resulting values of the power are comparable to the maximum power of Equation (\ref{Pj}) at $a_*\sim 1$ in several cases \citep{Carotenuto24, Cooper25}. The vast propagation distances require the presence of surrounding cavities with the densities $\ll 1$ cm$^{-3}$ \citep{Heinz02, SZ23}. Notably, such powerful ejections have been observed only among LMXBs. Thus,  some of them may be required to have high spins. On the other hand, \citet{Fender25} have compared the measured ejecta velocities with the published values of the spin obtained using the reflection, continuum, and QPO methods, and found no correlations. This shows that either the ejecta power is at most weakly dependent on the spin, or the published spin values are unreliable (which we advocate here), or both.

\begin{figure}[t!]
\centerline{\includegraphics[width=1.03\columnwidth]{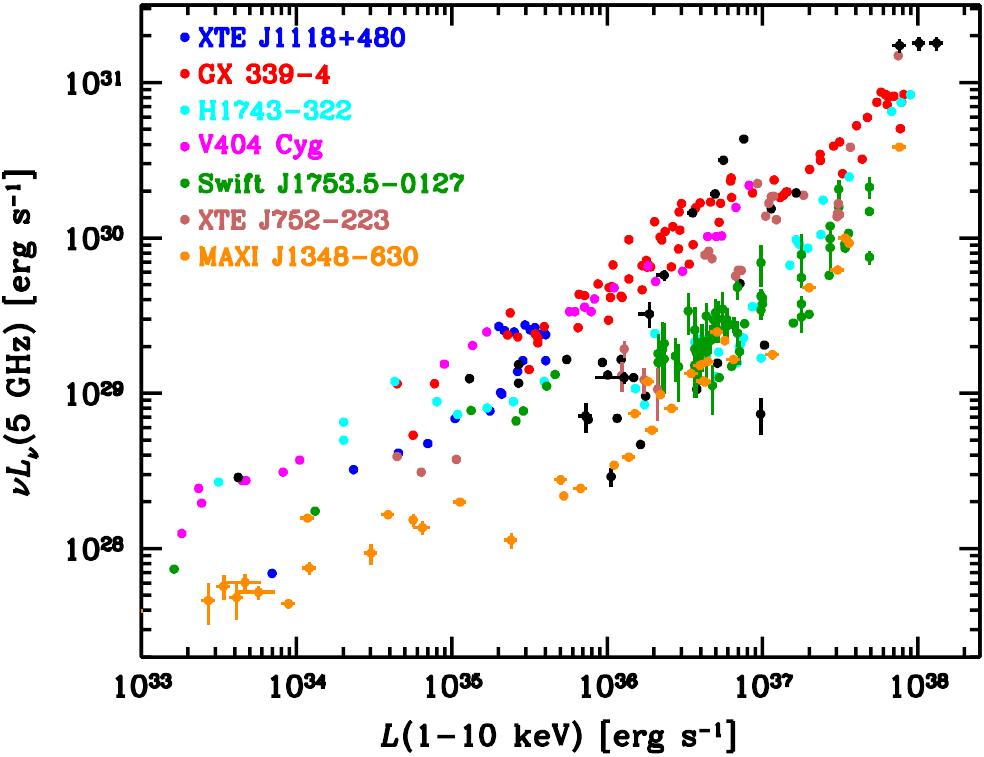}}
    \caption{The radio/X-ray correlation of the luminosities in BH LMXBs in the hard state, using the compilation by \citet{Bahramian22}. The black points show the data for MAXI J1659--152, GRO J1655--40, XTE J1720--318, IGR J17177--3656, MAXI J1836--194, GS 1354-64, and XTE J1650--500.
} 
\label{RX}
\end{figure}

On the other hand, the propagation distances of compact jets emitted in the hard state are much lower than those of transient ejections, even for the same source at very similar accretion rates, e.g., for MAXI J1820+070 \citep{Z_Heinz24}. Since both types of jets follow the same path in the surrounding medium, this implies that the power of compact jets is much lower than that of transient jets. If the compact jets are due to the spin extraction and obey Equation (\ref{Pj}), then the compact jets have $P_{\rm j}\ll\ \dot M_{\rm accr}c^2 a_*^2$, and their $P_{\rm j}$ may be independent of $a_*$. This can be tested using the correlation between the radio and X-ray luminosities in the hard state. We use here the compilation by \citet{Bahramian22}, shown in Fig.\ \ref{RX}. We see that the radio luminosities are spread by one order of magnitude or less for a given X-ray luminosity, with some of the spread due to a spread of the BH masses, different flux calibration of different X-ray instruments, and distance uncertainties. In addition, two main tracks appear to be present \citep{Coriat11}. Thus, the observed spread appears to be lower than expected if the jet emission were $\simprop a_*^2$, unless all sources have similar spins \citep{Fender10}. Additionally, we observe that the hard-state radio/X-ray luminosity ratios for MAXI J1348--630 were the lowest in the sample shown. At the same time, this binary featured a very powerful ejection observed to travel to $\sim$1 pc \citep{Carotenuto21}, which appeared to require launching from a MAD with $a_*\sim 1$ \citep{Zdziarski23a}. 

\citet{Daly19} used a method based on correlations between different bands to estimate the spins of unbeamed radio-loud AGNs as well as four BH LMXBs: GX 339--4, XTE J1118+480, V404 Cyg, and A0620--00. For these sources, the obtained spin values were $0.92\pm 0.06$, $0.66\pm 0.02$, $0.97\pm 0.02$, $0.98\pm 0.07$, respectively. Since these values were obtained from approximate and empirical scaling relations obtained for AGNs, they are strongly model-dependent. 

Then, we consider the jet power for the three known BH HMXBs with existing estimates of the spin. LMC X-1 and M33 X-3 have no reported radio jets. Cyg X-1 has a jet in its hard state \citet{Stirling01}, but its estimated power is low. The average hard-state luminosity assuming the distance of 2.2 kpc \citep{Miller-Jones21} is $\approx 2\times 10^{37}$ erg s$^{-1}$ \citep{ZPPW02, Wilms06}, implying $\dot M_{\rm accr}c^2\gtrsim 10^{38}$ erg s$^{-1}$. Based on its synchrotron radio emission, \citet{Heinz06} and \citet{Z_Egron22} estimated $P_{\rm j}\ll \dot M_{\rm accr}c^2$. Another constraint follows from assuming that a pc-scale structure discovered close to Cyg X-1 \citep{Gallo05} is powered by its approaching jet, which implies $P_{\rm j}\approx (1$--$3)\times 10^{37}$\,erg\,s$^{-1}$ \citep{Russell07}, which is still significantly below $\dot M_{\rm accr}c^2$. 

Finally, we mention the case of the Galactic super-Eddington source SS 433. It has powerful jets moving with a velocity of $0.26 c$ with the estimated kinetic luminosity of $\sim\! 10^{39}$ erg s$^{-1}$ (\citealt{Fabrika04} and references therein), i.e., $\approx L_{\rm Edd}$ for a $10\msun$ BH. While this is high, it is still below the Eddington accretion value, $\dot M_{\rm Edd}c^2$. Furthermore, the actual $\dot M_{\rm accr}$ is this system is most likely $> \dot M_{\rm Edd}$, see Equation (\ref{Mdot_ratio}) below. Thus, the jet power is $\ll \dot M_{\rm accr}c^2$ for all known BH HMXBs. 

\subsection{X-ray polarization and spins}
\label{polarization}

An independent method for measuring the BH spin is through X-ray polarization. The image of the innermost regions of the accretion disk is distorted by general and special relativity effects, causing different parts of the disk -- whose emission dominates at different energies -- to have varying orientations. This effect is expected to produce a gradual change of PA as a function of energy and progressive decrease of PD \citep{Dovciak04, Loktev22, Loktev24}. The rate of PA rotation and the suppression of PD are sensitive to BH spin, making them valuable diagnostic tools. However, X-ray polarization measurements of soft-state sources show neither statistically significant evidence of PA dependence on energy, nor do they indicate a decrease of PD with increasing energy \citep{Svoboda24, Marra24, Ratheesh24, Steiner24}.

The importance of returning radiation was proposed to explain the observed increase of PD with energy and the nearly constant PA in two sources, 4U 1957+115 and Cyg X-1 \citep{Marra24, Steiner24}. However, satisfying both criteria for the latter required an assumption of an extreme albedo of unity. For the former, the albedo was found to be not constrained. Moreover, no simultaneous fits of both the spectral and polarimetric signatures were done for the latter: either spectral modeling was performed separately from polarimetric modeling, or the parameters that provided a good fit for the polarization data failed to align with the spectral constraints. 

Returning radiation alone cannot account for the high PD observed in 4U 1630--47 \citep{Ratheesh24}. Furthermore, the high PD values fundamentally contradict predictions of the standard disk atmosphere models \citep{Chandrasekhar60, Sobolev63, Loskutov81}. Tight constraints on PA constancy with energy in this source placed strong upper limits on the BH spin, suggesting $a_*\leq 0.7$.

The absence of energy-dependent polarization signatures across both high- and low-inclination sources, the requirement for extreme albedo assumptions in the returning radiation scenario, and the failure of existing models to explain the polarization in 4U 1630--47 all hint towards an alternative mechanism for X-ray polarization production. One possibility is that the PD is enhanced by scattering in the disk wind launched at large distances from the central source \citep{Nitindala25}. Alternatively, the polarization may arise from Comptonization processes in a warm ($\sim10$~keV) disk atmosphere containing a fraction of non-thermal electrons (Bocharova et al., in preparation). However, in these scenarios, the constant PA with energy remains a viable probe of the BH spins, as the axisymmetric wind scattering does not introduce PA rotation.

\subsection{Published values of the BH spins of XRBs}
\label{published}

We know three BH HMXBs with measured spins: Cyg X-1, LMC X-1, and M33 X-7. Their spin values published before 2024 were all high: $>$0.9985 \citep{Zhao21_CygX1, Miller-Jones21}, $0.92^{+0.05}_{-0.07}$ \citep{Gou09}, and $0.84 \pm 0.05$ \citep{Liu08}, respectively. Those results were obtained using the disk continuum method in the soft state, specifically with {\tt kerrbb2}. A new value for Cyg X-1 was found in the spectral fit with {\tt kerrbb} by \citet{Steiner24}, $a_{*}= 0.99964^{+0.00003}_{-0.00007}$. Although those errors do not include systematic errors and uncertainties in the mass, distance, and inclination, very high values of the spin of Cyg X-1 were also reported using the previous determination of the mass and distance. In particular, \citet{Gou14} found $a_*>0.983$ using the values of $M=14.8\msun$ and $D=1.86$ kpc \citep{Orosz11}. 

Similarly, the published spin values for BH LMXBs are high on average, with $a_*\gtrsim 0.9$ in many cases and the mean value of $\langle a_*\rangle >0.7$; see the compilations in \citet{Reynolds21} and \citet{Fishbach22}. A recent paper by \citet{Draghis24} used the reflection method for a large sample of BH XRBs and claimed that the spins obtained in their work are even higher for most sources than those found previously. Namely, about 86\% of their sample had spins consistent with $a_*\geq 0.95$, 94\% were consistent with $a_*\geq 0.9$, and 100\% of their spin values were consistent with $a_*>0.7$. However, they fitted models with the incident spectrum being either an e-folded power law or a single thermal Comptonization spectrum. As we argued in Section \ref{reflection}, the actual incident spectra are significantly more complex. In particular, spectral fits in the hard state with two Comptonization components typically yield truncated disks (e.g., \citealt{Basak17, Zdziarski21b, Zdziarski21c, Chand24, Sahu25}), which then do not allow a spin measurement. On the other hand, \citet{Draghis24} assumed that the disk extends down to the ISCO in all cases, in both the hard and soft states. Taking their results at face value, even larger spins would be implied, with the obtained values being lower limits.

\begin{table}
\caption{Spin parameter estimates for 4U 1543--475 from various studies using reflection spectroscopy.}
\centering
\begin{tabular}{cc}
\hline
Reference & Spin parameter ($a_*$)\\
\hline
\citet{Miller09}  & 0.3 $\pm$ 0.1 \\
\citet{Morningstar_Miller14} & 0.43$^{+0.22}_{-0.31}$ \\
\citet{Dong20} & 0.67$^{+0.15}_{-0.08}$ \\
\citet{Draghis23} & 0.98$^{+0.01}_{-0.02}$\\
\citet{Draghis24} & 0.959$^{+0.031}_{-0.079}$\\
\citet{Yang_Jun24} & 0.902$^{+0.054}_{-0.053}$\\
\hline
\end{tabular}
\label{tab:spin4U1543}
\end{table}

An illuminating example is provided by the studies of the BH spin in 4U 1543--475. They produced conflicting results, with significant variations across the literature. Table \ref{tab:spin4U1543} summarizes the spin estimates for this source based on reflection spectroscopy alone (for continuum fitting, see \citealt{Shafee06, Yorgancioglu23}). The reported spin values vary widely, ranging from 0.3 to 0.5 \citep{Miller09, Morningstar_Miller14} to values exceeding 0.95 \citep{Draghis23, Draghis24}. These discrepancies are partly due to differences in the assumed disk inclination and other assumptions. For example, the first two estimates were obtained using the disk density of $10^{15}~\text{cm}^{-3}$ (the value used in {\tt relxill} before v.1.0), while the last two required an unphysically steep radial emissivity, $\simprop R^{-8}$, coupled with unrealistically high iron abundances, $>$7 times solar. Additionally, some fits are heavily dominated by reflection, e.g., with the reflection fraction{\footnote{Defined as the ratio of intensity emitted towards the disk compared to escaping to infinity, in the frame of the primary source, see \url{https://www.sternwarte.uni-erlangen.de/~dauser/research/relxill/.}}} of 5.4 for a NuSTAR observation (obsID 90702326006) in figure 10 in \citet{Draghis23}, despite the source being in the soft state. Other studies, such as \citet{Yang_Jun24}, require the reflection fractions to be as high as 10 when fitting with a variable disk density. These discrepancies, combined with the mentioned methodological caveats, suggest that different interpretations of the same data can arise depending on the given observation, model, and the fitting procedure used.

The case of 4U 1543--475 underscores how model uncertainties and differing assumptions can yield a wide range of spin estimates. Recent NICER and NuSTAR observations of this source, as discussed by \citet{Connors23}, have revealed "complex reflection characteristics that challenge even the most advanced reflection models available". This highlights the broader difficulty in obtaining robust spin measurements using reflection spectroscopy -- not only for this source but potentially for all systems. Given these challenges, reflection-based spin estimates should be especially treated with caution, as they may be affected by modeling issues stemming from the extreme complexity of the observed spectra. While current state-of-the-art models offer valuable insights, their limitations and potential for systematic errors must be carefully considered. We strongly advise researchers to thoroughly review prior work and consult model developers before engaging in reflection-based analyses.

Then, the spins have been measured to be low in some binaries while they showed powerful jets, apparently requiring high spins to power them (Equation \ref{Pj} in Section \ref{jet}). A prominent case is that of MAXI J1820+070. Using the continuum method, \citet{Guan21} and \citet{Zhao21} obtained $a_*= 0.2^{+0.2}_{-0.3}$ and $a_*= 0.14\pm 0.09$, respectively, while \citet{Fabian20} argued for $a_*<0$ based on the possible presence of emission from the plunging region below the ISCO (see Section \ref{continuum}). On the other hand, this system showed a powerful transient jet \citep{Bright20, Carotenuto24}, which requires $a_*\sim 1$ \citep{Z_Heinz24}.

\section{Potential solutions to the spin discrepancy}
\label{solutions}

There is thus a strong tension between the low spins inferred for the premerger BHs (especially for the first-born BHs) and the high spins determined from modeling the EM emission of BH XRBs (whose BHs are first-born). As confirmed by \citet{Fishbach22}, the distributions of $a_*$ of both BH HMXBs and BH LMXBs are incompatible with those inferred from mergers of BBHs. 

\subsection{Revised spin measurements and improved disk models}
\label{revised}

\begin{figure*}[t!]
\centerline{\includegraphics[width=8.5cm]{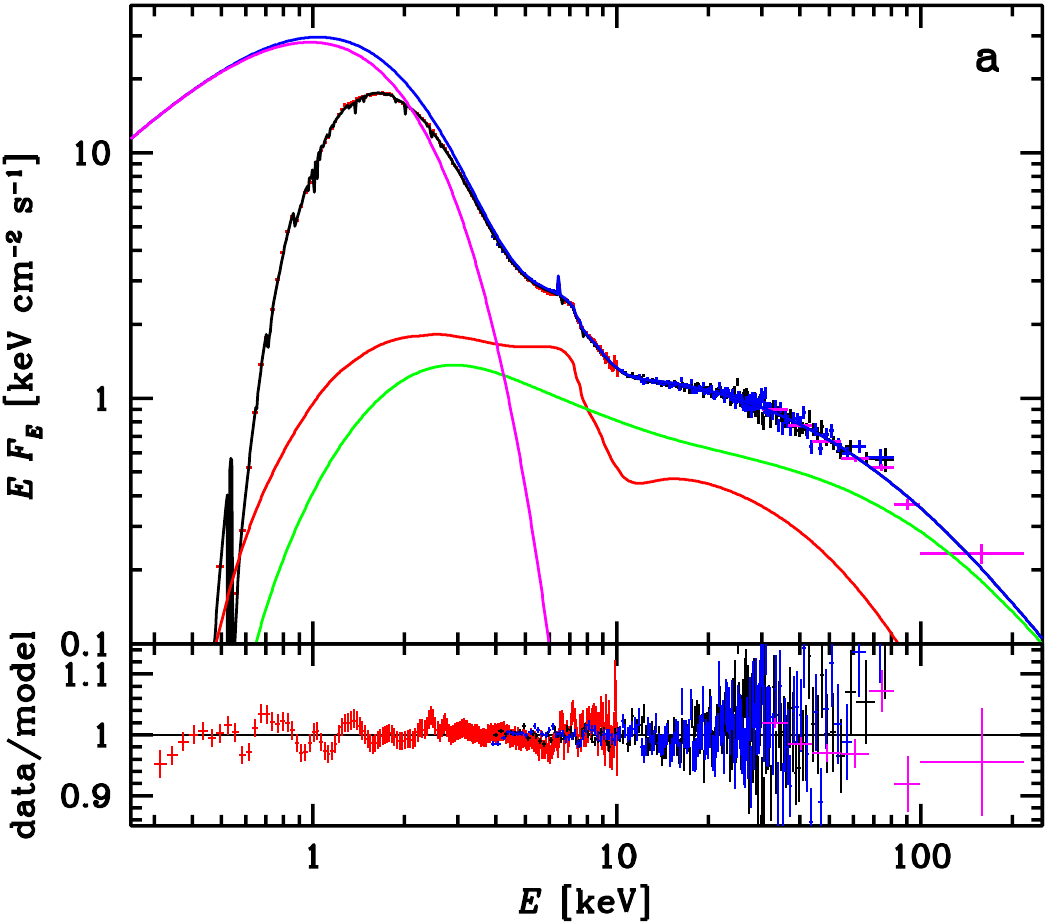} \includegraphics[width=8.5cm]{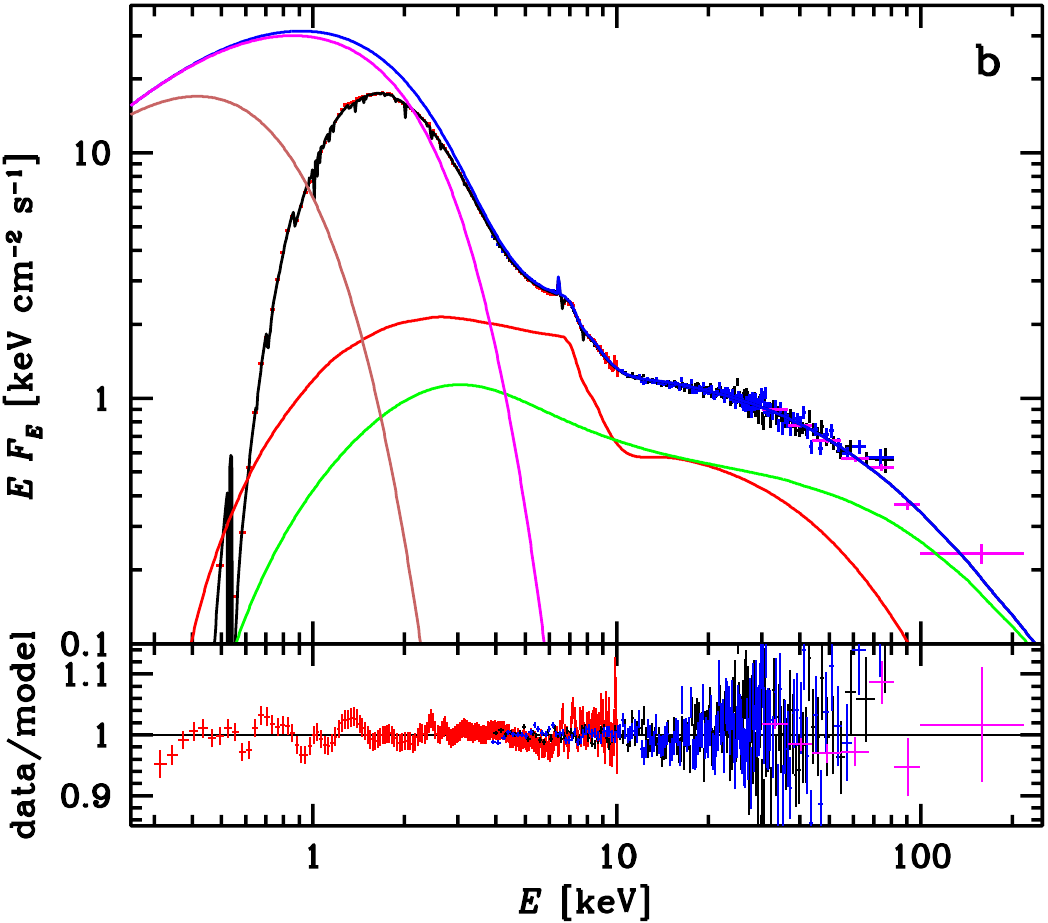}}
\caption{A soft-state state of Cyg X-1 observed by NICER, NuSTAR and INTEGRAL (based on \citealt{Zdziarski24b}). We show the unfolded spectra and data-to-model ratios in the top and bottom panels. In (a), the spectrum is fitted by the disk model {\tt kerrbb} and Comptonization, yielding $a_*=0.87^{+0.04}_{-0.03}$. In (b), the spectrum is fitted with a similar model but including a warm corona, and $a_*=0.00^{+0.07}$. The total model spectra and the unabsorbed ones are shown by the solid blue and blue curves, respectively. The unabsorbed disk, scattered, and reflected components are shown by the magenta, green, and red curves, respectively. In (b), the brown curve represents the underlying disk spectrum before it undergoes Comptonization in the top layer. See \citet{Zdziarski24b} for details.
}\label{cygx1}
\end{figure*}

In the case of GX 339--4, its mass and distance are only roughly constrained \citep{Heida17, Zdziarski19a}. Still, the continuum method, coupled with the reflection, has been applied to a study of this source by \citet{Parker16} and \citet{Zdziarski25a}. In the latter, the joint fitting of two sets of high-quality data in soft states with only weak high-energy tails from NICER and NuSTAR has allowed the authors to determine the BH spin. However, it turns out that the results strongly depend on the disk model used. The models treating departures from the local blackbodies by a color correction, in particular {\tt kerrbb} and {\tt kerrbb2}, yielded strongly negative spins. However, models employing radiative transfer calculations, namely {\tt bhspec} and {\tt slimbh}, yielded much better fits with moderately positive spins, and relatively high masses, in agreement with the mass function of \citet{Heida17}. The good fits of these two models show that the observed spectra agree much better with the atmospheric models than those employing color corrections. This is despite the standard disk model (assumed in the atmospheric models) disagreeing with the stability of disks in the soft state (Section \ref{continuum}). In particular, the disk stability is observed in these observations of GX 339--4 (with the rms variability of the disk emission $\lesssim$1\%). 

The study of \citet{Zdziarski25a} has also shown that even for fixed BH mass and distance, different available standard models (without including a warm corona) give significantly different spins; see Table 3 in \citet{Zdziarski25a}. In that example, the differences between the spins fitted by different disk-continuum codes are up to $\Delta a_*\approx 0.4$. Thus, the published results of fitting the standard continuum disk models should be considered significantly uncertain. 

Moreover, beyond the problems resulting from the choice of disk models, there are also issues regarding the choice of the source parameter, in particular, the color-correction factor. This problem has been addressed in \citet{Belczynski24} and \citet{Zdziarski24a, Zdziarski24b, Zdziarski25a} for the BH HMXBs Cyg X-1, LMC X-1, M33 X-7 and the LMXB GX 339--4. In the case of Cyg X-1, whose distance is well constrained and for the mass and inclination of \citet{Miller-Jones21}, the continuum fitting method with the standard method (at $f_{\rm col}=1.7$) and the binary viewing angle of $i=27.5^\circ$ yields results relatively similar to the previous studies, $a_*= 0.986^{+0.002}_{-0.001}$ \citep{Zdziarski24b}, while a somewhat lower spin, $a_*\approx 0.85$--0.90, is obtained for free $f_{\rm col}$ and $i$. On the other hand, we have checked that for the mass and inclination found as most likely by \citet{Ramachandran25}, $M\approx 13.8\msun$, $i=33.7^\circ$, and for $f_{\rm col}=1.7$, the spin is reduced from the previous value of $\approx$0.99 to $a_*=0.73\pm 0.01$.

Similar results are obtained for LMC X-1, with $a_*\approx 0.65$--0.96 using different variants of the standard model. In the case of M33 X-7, the previous determination of the BH mass of $\approx\! 15.6\msun$ \citep{Orosz07} has been revised to $\approx\! 11.4\msun$ \citep{Ramachandran22}, which resulted in a reduction of the spin from $a_*=0.84 \pm 0.05$ to $a_*\approx 0.7$ using a standard disk model. 

On the other hand, adding a warm corona in the soft state of Cyg X-1 reduces the spin dramatically, to $a_*\approx 0$--0.1 \citep{Belczynski24, Zdziarski24b}, see Fig.\ \ref{cygx1}. For GX 339--4, it yielded the spin only weakly constrained, $0\lesssim a_*\lesssim 0.8$, thus consistent with being very low. When adding a warm corona in LMC X-1, the spin becomes weakly constrained (due to the observed flux of this source being much weaker than that of Cyg X-1), with $a_*\approx 0.1$--0.9 \citep{Zdziarski24a}. Similarly, adding it in M33 X-7 leads to an almost unconstrained spin, $0\lesssim a_*\lesssim 0.9$ \citep{Belczynski24}. 

Thus, the recent studies of the spins including a warm corona in modeling have shown that in the cases studied so far (Cyg X-1, LMC X-1, M33 X-7, and GX 339--4) the BH spin is consistent with being low, though only in the case of Cyg X-1 the fit constrains it to $0\lesssim a_*\lesssim 0.1$. This offers a way to reconcile the spins from XRBs with those from mergers.

For the hard state, we consider the results of fitting with the reflection method to be unreliable due to the assumption adopted in those studies of the incident spectrum to be a single power law, while allowing more complex spectra (e.g., two Comptonization components) often yield truncated disks (which do not allow a spin measurement). In the soft state, this method can be coupled with the continuum method, making it more reliable, though with the caveats discussed above. 

\begin{figure*}[t!]
\centerline{\includegraphics[width=7cm]{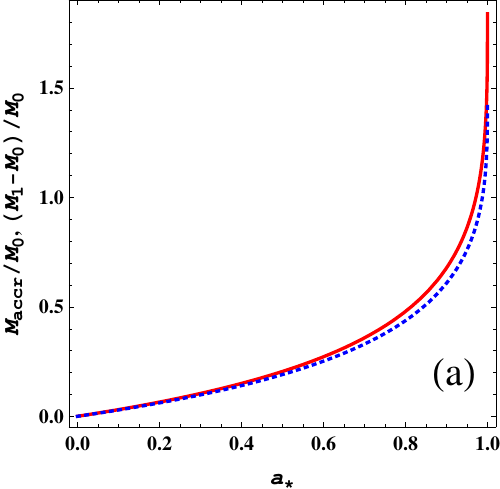}
\includegraphics[width=6.8cm]{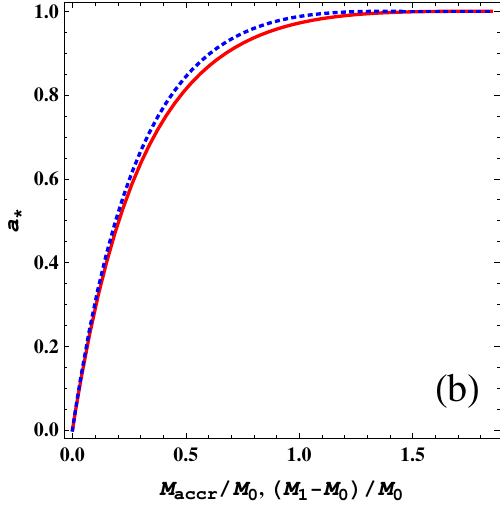}}
    \caption{\small (a) The red solid curve shows the fractional accreted mass required to reach a given spin, $a_*$, starting from null spin, and the blue dotted curve shows the corresponding fractional mass gained by the BH. The latter is lower than the former owing to the binding energy of the BH. The unit spin is achieved after $M_{\rm accr}/M_0=3\arccos\sqrt{2/3} \approx 1.846$, at which time $(M_1-M_0)/M_0 = 6^{1/2}-1\approx 1.449$. (b) The same, except that now the spin is shown as a function of the fractional mass accreted (red solid curve) and that gained by the BH (blue dotted curve).
} 
\label{DeltaM}
\end{figure*}

\subsection{Spin-up by accretion}
\label{accretion}

To spin up a BH with an initial mass $M_0$ from $a_*=0$ to $a_*=1$, a mass of $M_{\rm accr}=3 M_0 \arccos\sqrt{2/3} \approx 1.846 M_0$ needs to be accreted \citep{Bardeen70, Thorne74}. It is assumed that the matter accreted carries the specific angular momentum of the ISCO, and the decelerating power of jets \citep{Lowell2024, Lowell2025} is neglected. However, the BH mass grows by a lower amount, $M_1-M_0=(6^{1/2}-1)M_0\approx 1.449 M_0$, due to the rest energy being converted to the binding energy. Any additional accretion does not increase the spin. In terms of the final BH mass, the accreted fraction needed to spin up to $a_*=1$ is $M_{\rm accr}/M_1= \sqrt{2/3} \arccos\sqrt{2/3} \approx 0.754 $, while  $M_0/M_1=6^{-1/2} \approx 0.408$. For $M_1/M_0\leq 6^{1/2}$, a null initial spin and an arbitrary final spin, $a_*$, we have
\citep{Bardeen70, Thorne74}
\begin{align}
a_*&=\sqrt{\frac{2}{3}}\frac{M_0}{M_1}\left([4-\sqrt{ \frac{18 M_1^2}{M_0^2}-2}\right),
\label{spin_mratio}\\
\frac{M_1}{M_0}&=2 \sqrt{2}\sin\frac{M_{\rm accr}}{3 M_0}+\cos\frac{M_{\rm accr}}{3 M_0}, \label{mratio_maccr}\\
a_*&=\frac{\sqrt{\frac{2}{3}} \left\{4-\sqrt{\frac{18}{\left[2 \sqrt{2} \sin
   \left(\frac{ M_{\rm accr} }{3 M_0}\right)+\cos
   \left(\frac{ M_{\rm accr} }{3 M_0}\right)\right]^2}-2}\right\}}{2 \sqrt{2} \sin
   \left(\frac{ M_{\rm accr} }{3 M_0}\right)+\cos \left(\frac{ M_{\rm accr} }{3 M_0}\right)},\label{spin_maccr}\\
\frac{M_{\rm accr}}{M_0}&=3\left(\arcsin \frac{M_1}{3 M_0}-\arcsin\frac{1}{3}\right), \label{maccr_mratio}
\end{align}
where Equation (\ref{spin_maccr}) gives $a_*$ as a function of $M_{\rm accr}$, and Equation (\ref{maccr_mratio}) is the inverse function of Equation (\ref{mratio_maccr}). Fig.\ \ref{DeltaM}(a) shows the fractional accreted mass needed to spin up from 0 to $a_*$ and the corresponding (lower) mass gain by the BH, and Fig.\ \ref{DeltaM}(b) shows the spin vs.\ the accreted and gained masses. We see that the required accreted mass sharply decreases for $a_*<1$, e.g., $M_{\rm accr}\approx 0.677 M_0$ and $0.358 M_0$ for $a_*=0.9$ and 0.7, respectively. We see that a substantial mass is still needed to spin up a BH to a high $a_*$. If this explains the BH HMXBs' spins, we would then need to understand why most of the BBHs were not significantly spun up.

A non-zero misalignment slows down the spin-up. Equations (\ref{spin_mratio}) and (\ref{maccr_mratio}) were derived assuming the disk normal and the spin axis are aligned. In systems with a high natal kick, like the LMXB MAXI J1820+070 \citep{Atri19}, we expect a strong misalignment, which was confirmed by \citet{Poutanen22}. 

\subsubsection{BH HXMBs}
\label{HMXB}

Some BH HMXBs could be progenitors of binary BH mergers, so the need to explain the discrepancy is most immediate for them. In the studies of Cyg X-1, LMC X-1, and M33 X-7 of \citet{Zhao21_CygX1, Gou09, Liu08}, respectively, the BH spin is determined using the continuum method (Section \ref{continuum}). Those authors assumed that the high spins they measured were natal, which disagrees with the current views (Section \ref{GW}; \citealt{Fuller_Ma19}). The revised studies of those binaries of \citet{Belczynski24} and \citet{Zdziarski24a, Zdziarski24b} confirmed that their spins are still high (though somewhat lower than before) when fitted by the standard disk models. However, the short lifetimes of HMXBs prevent Eddington-limited accretion from substantially increasing the BH mass (and thus the spin). The Eddington time, i.e., the e-folding time for the mass increase, is independent of the BH mass,  
\begin{equation}
t_{\rm Edd}=\frac{\eta (1+X) c \sigma_{\rm T}}{8\pi (1-\eta) G m_{\rm p}},
\label{tedd}
\end{equation}
where $X$ is the H mass fraction. At $\eta\approx 0.1$, $X\approx 0.7$, it is $\approx$40\,Myr. For an increase of the mass corresponding to a spin increase from 0 to 0.9, $M_1/M_0\approx 1.68$ (see Fig.\ \ref{DeltaM} and the text above), the required time is $t_{\rm Edd}\ln (M_1/M_0)\approx 20$\,Myr. This is much longer than the typical lifetime of a BH HMXB, e.g., $\approx$4\,Myr for Cyg X-1 \citep{Miller-Jones21}. This conclusion was also achieved by a recent study of \citet{Xing25}, in which the conservative hypercritical accretion was confirmed (including the Roche-lobe overflow accretion channel) to be required to account for the measured spins of the three known BH HMXBs. Thus, we need hypercritical accretion (as it happens only episodically) if the high spins of those systems are real and their natal spins were low. 

Accretion rate onto a BH, $\dot M_{\rm accr}$, can exceed the Eddington rate under three circumstances. First, that limit can be overcome by photon advection into the BH (slim disks, e.g., \citealt{Abramowicz88}). Second, the disk can be geometrically thick, and the flow emission would then be strongly collimated into a narrow funnel, allowing an efficient mass flow onto the BH (Polish Doughnuts, e.g., \citealt{Jaroszynski80, Wielgus16}). Third, the energy and angular momentum of accreting matter can be efficiently transported away, in particular, by the emission of neutrinos. In those cases, the accretion rate onto the BH could, in principle, be equal to the mass transfer rate from the companion to the accretion disk, $\dot M_{\rm tr}$. 

However, this approach neglects outflows, which may be strong \citep{SS73}, causing a reduction of $\dot M_{\rm accr}$. \citet{Poutanen07} considered this problem in the context of ultraluminous X-ray sources. In their model, about half of the mass transferred from the donor crosses the horizon due to advection, i.e., $\dot M_{\rm accr}\sim 0.5\dot M_{\rm tr}$, which would be sufficient to explain the high spins. Similar values are obtained in numerical simulations for moderately super-Eddington mass transfer rates, up to $\dot M_{\rm tr}\sim 10 \dot M_{\rm Edd}$, see, e.g., \citet{Jiang14, Jiang19}, though \citet{Fragile25} found $\dot M_{\rm accr}\lesssim \dot M_{\rm Edd}$ in their simulations.

On the other hand, numerical simulations for $\dot M_{\rm accr}\gtrsim 10 \dot M_{\rm Edd}$ of \citet{Toyouchi24} (see their figure 10) yielded the (equivalent) relations, 
\begin{equation}
\frac{\dot M_{\rm accr}}{\dot M_{\rm tr}}\sim \left(\frac{\dot M_{\rm tr}}{\dot M_{\rm Edd
}}\right)^{-0.6}\!,\quad 
\frac{\dot M_{\rm accr}}{\dot M_{\rm Edd}}\sim \left(\frac{\dot M_{\rm tr}}{\dot M_{\rm Edd}}\right)^{0.4}\!,
\label{Mdot_ratio}
\end{equation}
where we conservatively assumed, following \citet{Toyouchi24}, that the termination radius, $R_{\rm t}$, below which the $\dot M$ no longer decreases with the radius, equals $6 R_{\rm g}$. If it is higher, $\dot M_{\rm accr}
$ increases $\propto\! (R_{\rm t}/6 R_{\rm g})^{0.6}$, see equation (37) of \citet{Toyouchi24}. This shows that while the accreted fraction becomes tiny at hyper-Eddington transfer rates, the accretion rate may still be super-Eddington, which is a very important result. If this result is confirmed, it should be taken into account in evolutionary models, where the accretion has so far been assumed as either Eddington-limited or conservative. 

We also note that slim disks/Polish Doughnuts transfer slightly lower/higher angular momenta to the BH than that of the Keplerian orbit at the ISCO, \citet{Sadowski_Narayan16} and \citet{Abramowicz80}, respectively. 

The need for hypercritical accretion for the BH HMXBs was also pointed out by \citet{Moreno08}, \citet{Moreno11}, and \citet{Qin22}. They considered energy loss from the accretion flow via the formation of neutrino pairs, which they claimed causes accretion to be fully conservative ($\dot M_{\rm accr}=\dot M_{\rm tr}$). This appears surprising; neutrino pair cooling is important in supernovae and GRBs, but the densities in hypercritical accretion flows are much lower, which will cause the flow to be optically thin to that pair production. Indeed, the simulation performed in \citet{Zdziarski24b} using the peak transfer rate for Cyg X-1 of the model of \citet{Qin22}, $\dot M_{\rm tr}=10^{-2}\msun$/yr, showed that the neutrino luminosity was only $\sim 10^{-5} \dot M_{\rm tr} c^2$, i.e., completely negligible. 

We have looked for the causes of this discrepancy. \citet{Moreno08}, \citet{Moreno11}, and \citet{Qin22} used the work by \citet{Brown94} to estimate the neutrino luminosity, specifically their equation (3.17), giving the emissivity of this process, $\dot \epsilon_{\rm n}$. \citet{Brown94} assumed in that equation that the electron chemical potential is fixed at $\mu=0$, in which case the average electron occupation number is close to unity and the formula for $\dot \epsilon_{\rm n}$ of equation (3.17) is independent of the electron density, $n_{\rm e}$. At $\mu=0$, the electron temperature equals the Fermi temperature,
\begin{equation}
T_{\rm F}=\frac{h^2}{8 k m_{\rm e}}\left(\frac{3 n_{\rm e}}{\pi}\right)^{2/3},
\label{Fermi}
\end{equation}
where $k$ is the Boltzmann constant, $m_{\rm e}$ is the electron mass and $h$ is the Planck constant. At $\dot M_{\rm tr}=10^{-2}\msun$/yr estimated by \citet{Qin22} for Cyg X-1, the average electron density within the spherization radius is $n_{\rm e}\sim 3\times 10^{14}$ cm$^{-3}$ (using the formalism of \citealt{Poutanen07}), which gives $T_{\rm F}\approx 0.2$ K. Obviously, the actual electron temperature of that flow is $T_{\rm e}\ggg T_{\rm F}$. Consequently, the gas is nondegenerate, whose chemical potential is negative,
\begin{equation}
\frac{\mu}{k T_{\rm e}}=\ln\frac{h^3 n_{\rm e}}{2 (2\pi m_{\rm e} k T_{\rm e})^{3/2}}
\label{mu}
\end{equation}
(e.g., \citealt{Cook95}). For example, for the above $n_{\rm e}$, and for $T_{\rm e}=10^6$ K, $\mu/kT_{\rm e}\approx -24$. Consequently, this invalidates the estimates of the neutrino luminosity in BH XRB accretion flows based on \citet{Brown94} and explains the discrepancy between the results of \citet{Zdziarski24b} and those of \citet{Moreno08}, \citet{Moreno11}, and \citet{Qin22}.

Summarizing this section, the main open questions for the claimed high spins of BH HMXBs are whether they could be natal and, if not, at which stage of the binary evolution the BH spins up via hypercritical accretion, and why those scenarios do not appear in BBHs. On the other hand, the actual HMXB spins could be low when fitted by alternative disk models, in particular those including warm coronae. 

\subsubsection{BH LXMBs}
\label{LMXB}

As stated in Section \ref{published}, the published values of the spins of BH LMXBs span the range from $\sim$0.1 to $>$0.99, with the mean value of $\langle a_*\rangle >0.7$. There are also some published claims of negative spins, for which the evidence appears relatively uncertain, with a detailed discussion provided in \citet{Zdziarski25a}. Assuming their BHs were born with low spin, they were spun up by accretion from the donor. In the first study of that process, \citet{King99} found that the Eddington-limited mass transfer rates and the theoretical lengths of the mass transfer rates in BH XRBs implied that the accretion could not significantly change the BH spins. This finding was revised in \citet{Podsiadlowski03}, who found a spin-up to $a_*\lesssim 0.9$ to be possible, provided the masses of the donors at the beginning of the mass transfer onto the BH were high, up to $M_{\rm 2,0}\sim 10\msun$. The difference concerning \citet{King99} was explained by taking into account both an increase of the Eddington rate at low spins (due to the low accretion efficiency) and an increase of the donor evolutionary lifetime as its mass decreases due to the mass transfer. Then, \citet{Fragos15} studied the same problem, but they assumed fully conservative accretion (i.e., not Eddington-limited). With this assumption, it was possible to obtain BHs with spins up to $a_*\approx 1$ and to reproduce all of the 16 Galactic BH XRBs with measured and estimated spins known at the time of their work. We note that the presence of conservative accretion is highly uncertain for phases with supercritical mass transfer, as discussed in Section \ref{HMXB}. \citet{Fragos15} also considered the case of Eddington-limited rates, for which their results were consistent with those of \citet{Podsiadlowski03}. Thus, they were unable to explain $a_*\gtrsim 0.9$, whereas $a_*\simeq 0.99$ has been claimed for some LMXBs. 

\begin{figure}[t!]
\centerline{\includegraphics[width=6.5cm]{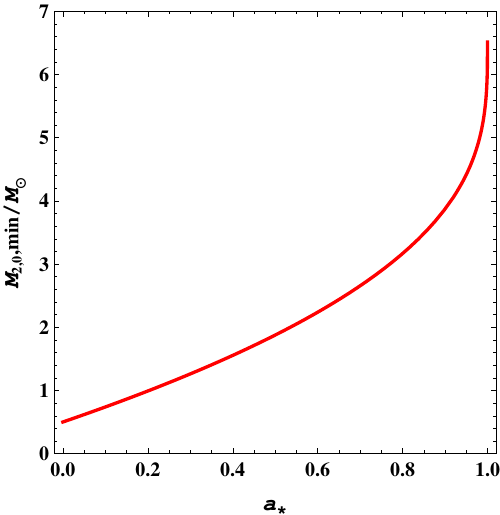}}
    \caption{\small An example of the dependence of the minimum donor mass at the onset of the Roche lobe overflow, $M_{\rm 2,0,min}$, required for a BH with the current mass of $M_1$ to be able to reach a given spin starting from null spin. Here, the current masses of $M_1=8\msun$ and $M_2=0.5\msun$ were assumed. This minimum corresponds to the accretion being fully conservative, and higher masses are required in the presence of outflows. 
} 
\label{minMd}
\end{figure}

The minimum (corresponding to the fully conservative accretion) donor mass at the onset of the mass transfer onto the BH, $M_{\rm 2,0,min}$, required to spin-up the BH from null spin to $a_*$ at the current binary masses $M_1$ and $M_2$ equals $M_{\rm accr}(a_*, M_1)+M_2$. For given $M_1$ and $a_*$, $M_0$ can be obtained by numerically solving  Equation (\ref{spin_mratio}), which can then be plugged into Equation (\ref{maccr_mratio}), which gives $M_{\rm accr}(M_0,M_1)$. This yields  
\begin{equation}
    M_{\rm 2,0,min}(a_*,M_1,M_2)= M_{\rm accr}[M_0(a_*,M_1),M_1]+M_2.
    \label{min_donor}
\end{equation}
An example is shown in Fig.\ \ref{minMd}. In it, $a_*=0.5$ yields $M_0\approx 5.3\msun$ and $M_{\rm 2,0,min}\approx 1.9\msun$. We also point out that spinning up a BH with a moderate mass to a high $a_*$ implies a small initial mass, which then can fall in the range of neutron-star masses. For example, the values of $M_1=6\msun$ and $a_*=0.998$ imply $M_0\approx 2.7\msun$. 

In a recent paper, \citet{Bartolomeo25} studied the accretion spin-up of the LMXB XTE J1550--564 using the spin measured by \citet{Steiner11} of $a_*=0.49^{+0.13}_{-0.20}$. They considered the measured values of $M_1$ and $M_2$ and the initial donor mass of $M_{2,0} \leq 1.4 \msun$. They studied several evolutionary scenarios starting from the onset of Roche-lobe overflow, concluding that the measured spin could be achieved only at its lower limit and for (unlikely) fully conservative accretion. We note, however, that this conclusion directly follows from Equation (\ref{min_donor}) {\it regardless\/} of the evolutionary details.

The initial donor masses in the models of \citet{Podsiadlowski03} and \citet{Fragos15} are often much higher than $\sim 1\msun$ typically observed in BH LMXBs. Those intermediate/high-mass donors will quickly lose most of their initial mass through mass loss, as the evolutionary time increases rapidly with decreasing mass. Thus, the probability of seeing an LXMB in the initial phases of mass transfer is very low. The initial mass distribution of companions in stellar binaries spans a wide range \citep{Sana2012}, although it may exhibit some dependencies on the initial orbital period \citep{Moe2017}. The initial orbital periods of the progenitors of LMXBs remain, however, poorly constrained due to the absence of a fully convincing mass-transfer scenario that can robustly reproduce the observed orbital separations of LMXBs within the framework of classical common-envelope evolution \citep[e.g.,][]{Kalogera1999, Podsiadlowski03, Wiktorowicz2014, Wang2016}. An additional potentially important ingredient is the high fraction of triple systems among massive stars \citep{Moe2017}. The presence of a tertiary companion may influence the early evolution of the inner binary through secular dynamical effects, such as the eccentric Kozai--Lidov mechanism \citep{Noaz2016}, thereby providing alternative pathways toward the formation of BH LMXBs.

Binaries with low-mass companions ($M_2\sim 1\msun$) may be more easily disrupted during BH formation, for example, due to the Blaauw kick associated with sudden mass loss \citep{Blaauw61}. This effect could reduce the survival probability of such systems and suppress their contribution to the observed BH LMXB population, thus favoring intermediate-mass companions as the progenitors of present-day BH LMXBs. On the other hand, another population of systems hosting BHs--including all three detached Gaia BH systems announced so far \citep{ElBadry2023a, Chakrabarti2023, ElBadry2023b, Gaia24}--contains companions with masses of the order of $\sim 1\msun$. These systems are found in the Galactic field and show no evidence for triple or higher-order multiplicity; their properties also pose a significant challenge for formation through conventional mass-transfer scenarios \citep{Nagarajan2025, Olejak2025}. Consequently, both observations and evolutionary models currently provide only weak constraints on the initial masses of the lower-mass companions in LMXB progenitors.

A potential way for testing the initial donor masses is through spectroscopic studies. \citet{Haswell02} studied the UV spectroscopy of XTE J1118+480. By considering the evolution of the system, they concluded that the donor mass at the onset of mass transfer was $M_{2,0}\approx 1.5\msun$. The currently estimated system masses are $M_1=7.6\pm 0.7\msun$ and $M_2=0.18\pm 0.07\msun$ (\citealt{Fragos15} and references therein), implying the accreted mass of $M_{\rm accr}=M_{2,0}-M_2\lesssim 1.4\msun$. Then, $M_2\ll M_{2,0}$ implies we see the exposed core of the system. This binary has no spin measurements. We can solve numerically Equation (\ref{maccr_mratio}), finding $M_0(M_{\rm accr},M_1)\geq 6.3^{+0.9}_{-0.7}\msun$, where the equality corresponds to conservative accretion. Then, the spin follows from Equation (\ref{spin_mratio}), yielding $a_*\leq 0.53^{+0.04}_{-0.05}$. 

The case of MAXI J1820+070 was studied by \citet{Georganti25}. They found that the ratios of the resonant emission lines N{\sc v}, C{\sc iv}, Si{\sc iv}, and the recombination line He{\sc i} are roughly consistent with those in CVs with standard abundances \citep{Mauche97}, see figure 14 in \citet{Georganti25}. This suggests that the initial donor mass was low enough to avoid CNO processing, $M_{2,0}\lesssim (1.0$--$1.5)\msun$. However, there are caveats to this picture. Namely, the emission line ratios depend on the ionization/excitation mechanisms present (in particular, collisional and/or photoionization are likely), the shape of the ionizing spectra, and abundances. The current system masses have been estimated as $M_1=6.75^{+0.64}_{-0.46}\msun$, $M_2= 0.49^{+0.10}_{-0.10}\msun$ \citep{Mikolajewska22}. Proceeding as above, we find $M_0\gtrsim 5.3\msun$ and $a_*\lesssim 0.51$. The spin upper limit is consistent with the low spin estimates of \citet{Guan21, Zhao21} based on the continuum method. On the other hand, this binary showed a powerful transient jet, which may require $a_*\sim 1$ (see Sections \ref{jet} and \ref{published}). Thus, more studies to resolve these conflicting pieces of evidence are needed. We also need studies comparing the known spectroscopic properties of the donors in BH LMXBs, many of which are classified as normal stars from dwarfs to giants (e.g., Table 1 in \citealt{Fragos15}), with the predictions of the stripped-donor model in cases with large initial donor masses. This would be a valuable test of the measured BH spin values.

\subsection{BBHs and BH HMXBs as two different binary populations}
\label{population}

The BBHs and BH HMXBs could, in principle, belong to two populations formed with two different evolutionary channels leading to low and high natal spins, respectively \citep{Fishbach22}. Using the data available in 2021, they found the possible existence of a subpopulation of binary BHs having the primary component rapidly spinning. However, this has not been confirmed for the sample of \citet{Abbott23}. 

\citet{Qin19} considered possible scenarios that can produce natal high-spinning BHs in HMXBs. They proposed a possible scenario with case A mass transfer within the BH progenitors while they are still on their main sequence. Obtaining very high BH spins, however, requires inefficient angular momentum transport, which is in tension with observations of stars as well as the GW population. As noted by \citet{Xing25}, this scenario is not supported by either observations or theory. Another way to obtain a natal high spin is through the impact of the supernova, but only up to $a_*\sim 0.8$ and under highly constrained circumstances \citep{Batta17, Moreno16, Schroder18}. Still, if the natal spins are high in BH HMXBs, it remains unclear why this is not the case for BBHs.

Binary population studies through the population synthesis method have shown that only a small fraction of BH HMXBs result in mergers \citep{Gallegos-Garcia22, Romero-Shaw23}. This still does not answer the question of the origin of the very high spins in BH HMXBs, see, e.g., discussion in \citet{Xing25}. 

In the case of the known BH HMXBs, the future evolution of Cyg X-1 will certainly lead to the formation of a BBH system, but the BHs will not merge in the Hubble time if the accretion is Eddington-limited. Only for fully conservative accretion, the system will merge within $\gtrsim$5 Gyr \citep{Ramachandran25}. M33 X-7 will, most likely, undergo a common-envelope stage during which the donor and the BH will merge; thus, it will not become a BBH \citep{Ramachandran22}. While the nature of the compact object in Cyg X-3 is not certain, it is most likely a BH \citep{Antokhin22}. However, its masses remain uncertain. If the mass of the WR donor is $\gtrsim 13\msun$, it will become a BBH within less than 1 Myr, and the BHs will most likely merge \citep{Belczynski13}. 

The BHs in mergers of field binaries also have undergone evolution in mass and spin, and in this respect are similar to XRBs. The crucial issue, which we stress in this paper, is that the GW results strongly imply that the natal spins are low.  To explain the spins measured in HMXBs, the natal spin has to be very high, in contrast to similar BHs undergoing mergers. No viable explanation has been proposed yet. For LMXBs, either the natal spins are high, which again is unexplained, or the donor masses at the onset of the RLOF were high, which remains to be shown, and which we discuss in detail. 

It has been proposed that failed supernova explosions can spin up BHs through the accretion of the remaining hydrogen envelope or fallback material. \cite{Antoni2022} considered accretion from the hydrogen envelope of a red supergiant progenitor and found that this mechanism does not produce spins higher than $a \simeq 0.8$. On the other hand, \cite{Batta17}, focusing on HMXBs, suggest that spins exceeding $a \gtrsim 0.8$ can be reproduced if a substantial amount of fallback material is generated during the failed explosion, and if its interaction with the companion star increases its angular momentum content. Once this material is eventually accreted, it may lead to a rapidly spinning BH. We argue, however, that the failed supernova spin-up mechanism is unlikely to resolve the apparent spin tension between the GW and HMXB populations, and may be challenged for several reasons.

In particular, at least a fraction of BBH merger events---especially those involving very massive BHs \citep{Fryer12}---are also expected to form through failed supernovae or direct collapse. The associated low natal kicks in such explosions would help binaries avoid disruption and merge within a Hubble time. Yet current GW observations do not show evidence for high spins, and certainly not for spins approaching $a \sim 0.8$. Moreover, if failed supernovae were the dominant pathway to producing highly spinning BHs, one would expect a correlation between BH mass and final spin above some threshold for direct-collapse formation \citep{Fryer12}. While the most massive BHs in the current BBH sample (with masses $\gtrsim 30\,\msun$) may show hints of moderately elevated spins \citep{LVK25a}, these masses are significantly higher than those inferred for X-ray binaries. BHs detected via GWs with masses comparable to typical XRB BHs ($\lesssim 20\,\msun$) instead remain consistent with $a_{\rm eff} \sim 0$ and low individual spins.

Finally, as discussed above, BH progenitors in both populations undergo mass-transfer phases that likely remove a substantial fraction of the hydrogen envelope. In addition, XRBs are frequently found in relatively high-metallicity environments, where stellar winds can remove up to $\sim 80\%$ of the progenitor's initial mass \citep{Belczynski2010}, carrying away a significant amount of angular momentum. If the failed supernova spin-up scenario were operating efficiently, one would naively expect GW sources---often forming in lower-metallicity environments, experiencing weaker stellar winds (and thus more likely retaining part of their hydrogen envelopes)---to host more rapidly spinning BHs than those observed in XRBs. This expectation is contrary to current observations.

\section{Discussion and summary}
\label{discussion}

We summarize the main points of tension between the BH spins obtained in BH X-ray binaries and those from the GW events, critically assess potential obstacles in spin determination methods, and list potential ways to resolve systematic biases in these studies.

(i) {\it Natal spins of BHs in merger events are found to be low.}
As studies of GWs have found, BBHs are characterized by low spins on average. This is especially the case for the first-born BHs; see Fig.\ \ref{fig17}. This also implies that the natal spins of first-born BHs are low.

(ii) {\it Lack of high spins in premerger BHs are in the strongest tension with the results from EM observations of BH HMXBs.} Some of them can be progenitors of BBHs, whereas the spins of the three known BH HMXBs are all high when using standard disk models, $a_*\sim 1$. If the high spin is natal (which is not supported by either theory or observations), then we have the question of why they are not high in first-born premerger BHs. Then, their short lifetimes prevent substantial spin-up by accretion, unless for hyper-Eddington rates, which are not predicted theoretically. If BH HMXBs still undergo such accretion, then we again have the question of why such accretion does not occur in progenitors of BBHs. 

(iii) {\it The spins of BH HMXBs could be low, due to the models giving high spins being inadequate.} The existing methods of measuring the spins are all highly uncertain, as we show in detail in Section \ref{EM}. The X-ray reflection method strongly depends on the decomposition of the incident and reflected spectra, which remain weakly determined. It also requires that the disk extends to the ISCO or very close to it, which is the assumption being challenged by spectral and timing data in the hard spectral state of XRBs. At the same time, all variants of the continuum-fitting method rely on the model of \citet{SS73, NT73}, which predicts the disk to be unstable; however, the observations show very stable disks, thereby disproving the validity of this method as well. A stable alternative is provided by models of strongly magnetized disks, which give higher atmospheric temperatures and thus imply significantly lower spins. Thus, our hypothesis regarding BH HMXBs is that their spins are $a_*\ll 1$, in agreement with the GW results. 

(iv) {\it Some BH LMXBs could have high spins if their initial donor masses were intermediate to high.} The currently measured distribution of spins for BH LMXBs covers the range approximately from 0 to 1, but the average value of the spin is high, and there are several cases with the spin of $a_*\approx 1$. From both theory and observations, their natal spins should be low. The high spins could be acquired by accretion, but only if the initial mass of their donors was at least several solar masses. This remains uncertain given the appearance of the observed donors compatible with the predictions of standard stellar evolution of LMXBs, where the initial masses are $M_{\rm d}\lesssim 2\msun$ and yield $a_*\lesssim 0.5$, where the upper limit corresponds to conservative accretion. Then, the hypothesis that most donors in BH LMXBs had high masses conflicts with the known steepness of the initial stellar mass function, which would predict a dominance of systems with low initial masses, and thus low spins. 

(v) {\it High spins of some BH LXMBs appear to be required by the observations of powerful transient jet ejections.}
Some BH LMXBs show jets of power comparable to their $\dot M_{\rm accr}c^2$ (Section \ref{jet}), implying their spins are of $a_*\sim 1$. 
At the same time, the jets can efficiently extract power from the BH and significantly reduce their spins. 
Thus, our conclusion for this class of systems is that their true spins remain uncertain but are mostly low, in agreement with the GW results, although there are some systems with high spins achieved via accretion from their donors with initial high masses. 

(vi) {\it New, realistic models of accretion disks are ultimately needed to extract BH spins from spectral fitting.}
We propose that the best way to reconcile the GW and EM spin results is through developing and using realistic accretion disk models. Such models should be stable in the regime of $0.05\lesssim L/L_{\rm Edd}\lesssim 1$, which can be achieved if a large part of the pressure support is provided by magnetic fields. Such disks are geometrically thicker than the standard disks and have dissipation in their surface layers. Such dissipation occurs, in particular, in warm coronae above the disks, which were shown to explain the broadband spectra of AGNs. An addition of such a corona yields $a_*\sim 0$ in Cyg X-1 and values compatible with those for LMC X-1, M33 X-7, and GX 339--4.

\section*{Acknowledgements}
We thank Sudeb Datta, Shane Davis, Gulab Dewangan, Julian Krolik, Piero Madau, Joanna Miko{\l}ajewska, Ranjeev Misra, Andrew Mummery, Andrzej Nied{\'z}wiecki, and Thomas Tauris for their comments. We also thank the two referees for their thoughtful suggestions. We acknowledge support from the Polish National Science Center grants 2019/35/B/ST9/03944 and 2023/48/Q/ST9/00138. Alexandra Olejak acknowledges funding from the Netherlands Organization for Scientific Research (NWO) as part of the Vidi research program BinWaves (project number 639.042.728, PI: de Mink). Gr{\'e}goire Marcel and Alexandra Veledina acknowledge support from the Academy of Finland grant 355672. Nordita is supported in part by NordForsk. Debora Lan\v{c}ov\'{a} acknowledges the Czech Science Foundation (GA\v{C}R) project No. 25-16928O.

  \newcommand\aap{A\&A}                
\let\astap=\aap                          
\newcommand\aapr{A\&ARv}             
\newcommand\aaps{A\&AS}              
\newcommand\actaa{Acta Astron.}      
\newcommand\afz{Afz}                 
\newcommand\aj{
AJ}                   
\newcommand\ao{Appl. Opt.}           
\let\applopt=\ao                         
\newcommand\aplett
{Astrophys.~Lett.} 
\newcommand\apj{ApJ}                 
\newcommand\apjl{ApJL}                
\let\apjlett=\apjl                       
\newcommand\apjs{ApJS}               
\let\apjsupp=\apjs                       
\newcommand\apss{Ap\&SS}             
\newcommand\araa{ARA\&A}             
\newcommand\arep{Astron. Rep.}       
\newcommand\aspc{ASP Conf. Ser.}     
\newcommand\azh{Azh}                 
\newcommand\baas{BAAS}               
\newcommand\bac{Bull. Astron. Inst. Czechoslovakia} 
\newcommand\bain{Bull. Astron. Inst. Netherlands} 
\newcommand\caa{Chinese Astron. Astrophys.} 
\newcommand\cjaa{Chinese J.~Astron. Astrophys.} 
\newcommand\fcp{Fundamentals Cosmic Phys.}  
\newcommand\gca{Geochimica Cosmochimica Acta}   
\newcommand\grl{Geophys. Res. Lett.} 
\newcommand\iaucirc{IAU~Circ.}       
\newcommand\icarus{Icarus}           
\newcommand
\japa{J.~Astrophys. Astron.} 
\newcommand\jcap{J.~Cosmology Astropart. Phys.} 
\newcommand\jcp{J.~Chem.~Phys.}      

\newcommand\jgr{J.~Geophys.~Res.}    
\newcommand\jqsrt{J.~Quant. Spectrosc. Radiative Transfer} 
\newcommand\jrasc{J.~R.~Astron. Soc. Canada} 
\newcommand\memras{Mem.~RAS}         
\newcommand\memsai{Mem. Soc. Astron. Italiana} 
\newcommand\mnassa{MNASSA}           

\newcommand\mnras{MNRAS}             
\newcommand\na{New~Astron.}          
\newcommand\nar{New~Astron.~Rev.}    
\newcommand\nat{Nature}              
\newcommand\nphysa{Nuclear Phys.~A}  
\newcommand\pra{Phys. Rev.~A}        
\newcommand
\prb{Phys. Rev.~B}        
\newcommand\prc{Phys. Rev.~C}        
\newcommand\prd{Phys. Rev.~D}        
\newcommand\pre{Phys. Rev.~E}        
\newcommand\prl{Phys. Rev.~Lett.}    
\newcommand\pasa{Publ. Astron. Soc. Australia}  
\newcommand\pasp{PASP}               
\newcommand\pasj{PASJ}               
\newcommand\physrep{Phys.~Rep.}      
\newcommand\physscr{Phys.~Scr.}      
\newcommand\planss{Planet. Space~Sci.} 
\newcommand\procspie{Proc.~SPIE}     
\newcommand\rmxaa{Rev. Mex. Astron. Astrofis.} 
\newcommand\qjras{QJRAS}             
\newcommand\sci{Science}             
\newcommand\skytel{Sky \& Telesc.}   
\newcommand\solphys{Sol.~Phys.}      
\newcommand\sovast{Soviet~Ast.}      
\newcommand\ssr{Space Sci. Rev.}     
\newcommand\zap{Z.~Astrophys.}       

\footnotesize
\bibliographystyle{elsarticle-harv} 
\bibliography{../allbib.bib}

\end{document}